\RequirePackage{etoolbox}

\newcommand{\setversion}[1]{\def\paperVersion{#1}}
\newcommand{\version}[2][]{%
    \ifdefstring{\paperVersion}{#1}{#2}{}%
}

\setversion{A} 

\version[A]{%
    \documentclass{article}
    \usepackage{arxiv}
  
    \renewcommand{\headeright}{}
    \renewcommand{\undertitle}{}
}
\version[B]{%
    \documentclass[letterpaper]{article}
    \usepackage[letterpaper,margin=1in]{geometry}
}

\usepackage{graphicx} 
\usepackage{orcidlink}
\usepackage{authblk}

\usepackage{amssymb} 
\usepackage{xcolor}

\usepackage{siunitx}
\usepackage{physics}
\AtBeginDocument{\RenewCommandCopy\qty\SI}
\ExplSyntaxOn
\msg_redirect_name:nnn { siunitx } { physics-pkg } { none }
\ExplSyntaxOff

\usepackage[%
style=base,
singlelinecheck=false,%
format=plain,%
labelformat=simple,%
labelsep=period,%
justification=justified,%
font=none,small,%
labelfont=bf,%
textfont=none%
]{caption}
\usepackage{subcaption}
\usepackage{cleveref}
\crefname{figure}{figure}{figures}
\Crefname{figure}{Figure}{Figures}
\crefname{table}{table}{tables}
\Crefname{table}{Table}{Tables}
\usepackage{csquotes}
\usepackage[
backend=biber,
style=numeric-comp,
sorting=none,
sortcites=true
]{biblatex}
\usepackage{hyperref}

\makeatletter
\def\@email#1#2{%
 \endgroup
 \patchcmd{\titleblock@produce}
  {\frontmatter@RRAPformat}
  {\frontmatter@RRAPformat{\produce@RRAP{*#1\href{mailto:#2}{#2}}}\frontmatter@RRAPformat}
  {}{}
}%
\makeatother

\newcommand{\Table}[1]{\begin{table*}
  \setlength{\tabcolsep}{8pt}
  \caption{#1}
  \begin{tabular}{@{}l*{15}{l}}}

\def\endTable{\end{tabular}\end{table*}}

\title{Fading Memory and Nonlinear Dynamics in a Microring Resonator Network and their Use in Computation}

\date{\today}

\author[1,2]{\orcidlink{0009-0001-3195-8620} Alessandro Foradori}
\author[2]{\orcidlink{0000-0002-6587-2614} Alessio Lugnan}
\author[2]{\orcidlink{0000-0001-7316-6034} Lorenzo Pavesi}
\author[1]{\orcidlink{0000-0001-6259-464X} Peter Bienstman}

\affil[1]{Photonics Research Group, Department of Information Technology, Ghent University-imec, Technologiepark Zwijnaarde 15, Ghent, 9052, Belgium}
\affil[2]{Nanoscience Laboratory, Department of Physics, University of Trento, Via Sommarive 14, Trento, 38123, Italy}

\begin{document}

\maketitle

\begin{abstract}
Silicon microring resonators are promising components for photonic reservoir computing because their nonlinear optical response can transform signals and retain information about recent inputs.
Under constant optical excitation, their coupled free-carrier and thermal dynamics can generate self-pulsing: a periodic output response.
However, constant-input behavior does not reveal how these dynamics respond to randomly modulated signals or whether their memory is useful for temporal processing.
Here, we characterize a single silicon microring resonator and a passive network of microring resonators under random input modulation.
The temporal component associated with self-pulsing remains observable under random modulation, but the largest fading memory occurs near the self-pulsing boundary, where no resolved self-pulsing-like oscillation is present.
reservoir computing metrics, such as memory capacity, map this accessible memory, and its decay provides effective timescales associated mainly with free-carrier and thermal relaxation.
Finally, we use NARMA reconstruction to test whether the measured memory profile supports nonlinear temporal processing.
Responses with a resolved self-pulsing-like component perform worse on this task, and the intrinsic memory profile is not matched to the required input history.
These results show that reservoir computing metrics can characterize nonlinear microring resonator dynamics even when the physical system is not an effective standalone reservoir for the selected task.
\end{abstract}

\section{Introduction}\label{sec:introduction}

Neuromorphic photonics uses optical hardware for neural-network operations and dynamical information processing \cite{shastri2021photonics}.
Photonic systems can provide high bandwidth, low propagation latency, energy-efficient linear transformations, and parallelism across wavelength and spatial channels \cite{shastri2021photonics,li2024photonics}.
These properties support time-dependent signal processing in optical communications and sensing.
On the other hand, processing such signals in the optical domain can reduce signal conversion and data movement \cite{argyris2022photonic,foradori2026memoryreview}.
Scalable neuromorphic photonics requires integrated nonlinear transformations and memory.
These functions are generally more difficult to implement optically than passive linear operations \cite{farmakidis2024integrated,foradori2026memoryreview}.

Here, we investigate an alternative, namely silicon microring resonators (MRRs), which are compact devices that combine wavelength-selective filtering, optical nonlinearity, and memory \cite{biasi2024photonic}.
At low intracavity optical power, usually $\ll \qty{1}{\milli\watt}$, an MRR acts as a wavelength-selective filter, and resonant recirculation enhances the interaction of light with silicon \cite{bogaerts2011silicon}.
At higher intracavity power, $\gtrsim \qty{1}{\milli\watt}$, two-photon absorption generates free carriers and contributes to heating.
Free-carrier-dispersion and thermo-optic effects then shift the resonance in opposite directions \cite{borghi2021the}.
The carrier and thermal responses relax on different timescales, and their coupled dynamics provide nonlinear transformations and fading memory \cite{borghi2021the,biasi2024photonic}.
Suitable input powers and detunings between the laser frequency and the MRR resonance frequency can produce self-pulsing under constant excitation.
Self-pulsing is a sustained temporal oscillation in which carrier- and temperature-induced resonance shifts repeatedly modulate the intracavity and output optical power \cite{vanvaerenbergh2012simplified,biasi2024exploring}.
The output response of a single MRR therefore depends on the laser frequency and input optical power.
MRRs also serve as tunable linear weighting elements in MRR weight banks \cite{tait2016microring}.
Other implementations use nonlinear MRR dynamics to realize all-optical silicon spiking neurons \cite{xiang2022all,donati2025all}.
Numerical studies examine how silicon MRR nonlinearities and losses affect time-delay reservoir computing and analyze or engineer memory capacity in single MRRs and series-coupled MRR systems \cite{gironcastro2024effects,ren2024photonic}.

An MRR network connects multiple resonators through bus waveguides that provide multiple optical propagation paths from input to output.
The MRR network response also depends on the output port because each port samples a different combination of resonances and propagation paths.
Measurements at different laser frequencies and output ports therefore provide different observations of the same underlying MRR network dynamics \cite{mancinelli2014chaotic,foradori2026timeseries}.
MRR networks have been applied to memory and time-series processing \cite{biasi2024photonic,lugnan2025nonfading,foradori2026timeseries}.

Reservoir computing (RC) provides a framework for evaluating these dynamical responses.
This framework treats a physical system as a dynamical processor and trains only a linear readout of its measured state features to reconstruct a target function of the time-dependent input \cite{jaeger2004harnessing,maass2002real}.
The general RC methodology and its photonic implementations are reviewed in \cite{lukosevicius2009reservoir,vandersande2017advances,tanaka2019recent}.
RC relies on fading memory, nonlinear transformation, and access to sufficiently diverse measured features \cite{dambre2012information,inubushi2017reservoir}.
Performance on machine-learning benchmark tasks reflects diverse computational requirements and depends on both the reservoir dynamics and the representation supplied to the readout, including its dimension, selected features, and any explicitly delayed inputs.
Task performance alone therefore cannot determine how much fading memory or nonlinear transformation originates in the photonic response.
By contrast, memory capacity (MC) directly estimates a reservoir's fading memory by measuring how accurately an optimized linear readout of the selected features reconstructs previous inputs.
Information-processing capacity (IPC) applies the same reconstruction test to nonlinear functions of delayed inputs \cite{dambre2012information}.
These metrics characterize a physical reservoir through its observable input--output responses without requiring access to every internal dynamical state.

Our previous studies of MRR-based photonic networks address different physical systems and operating regimes.
One study investigates a network containing a nonvolatile GST phase-change element \cite{lugnan2024emergent}.
Its architecture and memory mechanism differ from those of the fully passive carrier--thermal MRR network considered here.
Another study uses the same passive MRR network architecture as the present work and reports non-fading memory over tens of microseconds \cite{lugnan2025nonfading}.
It encodes information through temporary perturbations of a constant optical drive and reads the resulting self-pulsing response after the perturbations end.
A self-pulsing attractor is a stable periodic evolution of the network state, sustained by the constant optical drive.
A perturbation can transfer the system between coexisting attractors or shift the temporal phase of an oscillation, as illustrated experimentally in coupled MRRs \cite{biasi2024exploring}.
These mechanisms are distinct: attractor selection stores information in the identity of the retained dynamical state, but phase encoding stores information in the timing of the oscillation.
The state-selection mechanism is conceptually related to attractor-based recurrent neural networks, which encode and retain information in stable states supported by their recurrent dynamics \cite{hopfield1982neural,khona2022attractor}.
However, the complete attractor landscape of the MRR network has not been experimentally reconstructed, and a retained phase shift alone does not establish switching between distinct attractors \cite{foradori2026memoryreview}.
Although the previous study uses reservoir computing readout methods, its persistent-state operation differs from the fading-memory processing of continuously modulated inputs investigated here.
A third study uses the self-pulsing dynamics of the same architecture to transform MNIST and Fashion-MNIST images serialized as time sequences, and selects responses across output ports, laser frequencies, and input powers for classification tasks \cite{foradori2026timeseries}.
A complete serialized image as a time sequence in that study lasts approximately \qtylist{1.6;8;16}{\micro\second}, shorter than or comparable to the observed self-pulsing periods of approximately \qtyrange{1}{50}{\micro\second}.

A central question for nonlinear MRR networks is how their free-carrier--thermal dynamics retain and transform continuously modulated inputs.
Constant-excitation measurements reveal self-pulsing, but they do not show whether the associated temporal oscillation remains observable under random input modulation \cite{lugnan2025nonfading}.
Task-level performance also does not identify the effective physical relaxation scales or nonlinear transformations that support processing.
MC and IPC address these questions by quantifying fading memory and delayed nonlinear transformations from the measured input--output responses.

This work characterizes a single MRR and a 64-MRR network under randomly modulated excitation.
Temporal spectra compare the constant- and random-excitation dynamics, MC maps accessible fading memory and its decay, and IPC resolves nonlinear transformations of delayed inputs.
NARMA-$N$ reconstruction then tests whether memory and interactions among input symbols support a recurrent temporal task \cite{atiya2000new,wringe2025reservoir}.
These analyses lead to the following findings:
\begin{itemize}
	\item The self-pulsing-associated output-oscillation component persists under random modulation, but the largest MC values occur near the constant-drive self-pulsing boundary rather than where the resolved oscillation is strongest.
	\item RC metrics provide an input--output method for characterizing the physical system.
	The MC landscape identifies where input history is accessible, and the MC decay provides effective fast and slow timescales associated with free-carrier and thermal relaxation.
	\item The intrinsic memory profile of the passive MRR network is not sufficient for the evaluated NARMA temporal sequence-emulation tasks under the investigated conditions.
	Increasing the maximum symbol depth through digital storage supplies the missing history.
\end{itemize}

\Cref{sec:operating-principle} explains how the photonic systems, random excitation, measured-response diversity, and staged characterization support these analyses.
\Cref{sec:results} presents the spectral analysis, memory-capacity characterization, information-processing-capacity analysis, and NARMA reconstruction results.
The \nameref{sec:conclusion} (\cref{sec:conclusion}) summarizes the findings and future directions.
 section 4
\Cref{sec:methods} describes the photonic circuit, experimental setup, data acquisition, and analysis procedures.

\section{Evaluation Methodology}\label{sec:operating-principle}

This section outlines the methodology used to characterize how free-carrier--thermal MRR dynamics retain and transform continuously modulated inputs.
We introduce a single MRR as a baseline and then an MRR network, whose multiple resonances, propagation paths, and output ports provide diverse measured responses.
For each random input realization, we replay the same programmed sequence when changing the laser frequency, mean input power, output port, or device.
This common input allows separately measured responses to be compared and combined at matched sequence positions.
Finally, we use a sequence of complementary analyses to connect observable dynamics to fading memory, nonlinear transformation, effective relaxation scales, and NARMA processing.

\subsection{Photonic Systems and Nonlinear Dynamics}\label{sec:photonic-systems}

At sufficiently high intracavity powers, free-carrier and thermal effects changes the nonlinear refractive index which in turns shifts the MRR resonance respectively towards the blue and the red on different relaxation timescales.
The coupled response of these effects can therefore transform the input nonlinearly and retain information about its recent history.
The competing resonance shifts can produce a periodic output response to a constant excitation named self-pulsing \cite{borghi2021the,vanvaerenbergh2012simplified}.

\begin{figure}[htp]
	\includegraphics[width=\linewidth]{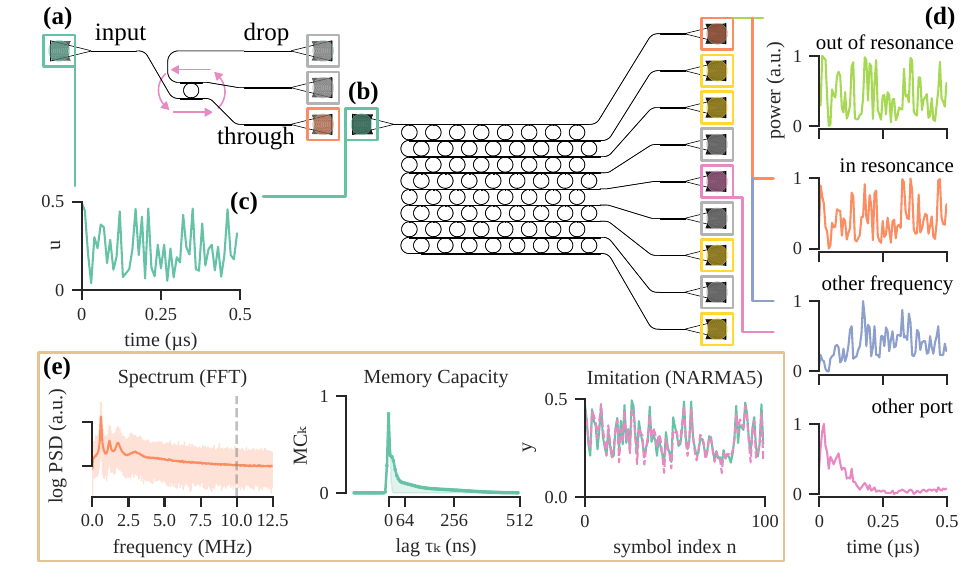}

	\begin{subcaptiongroup}
		\phantomcaption\label{fig:intro-ring}
		\phantomcaption\label{fig:intro-matrix}
		\phantomcaption\label{fig:intro-input}
		\phantomcaption\label{fig:intro-responses}
		\phantomcaption\label{fig:intro-metrics}
	\end{subcaptiongroup}

	\caption{%
		\textbf{Photonic systems, measured responses, and selected characterization analyses.}
		\subref{fig:intro-ring} Single MRR; aqua and orange frames identify the input and through ports respectively, and the pink arrows indicate optical circulation in the resonator.
		\subref{fig:intro-matrix} Passive $8\times8$ network of 64 MRRs with one input and nine output ports, numbered from top to bottom.
		Orange, yellow, and pink frames identify the six measured output ports (1, 2, 3, 5, 7, and 9); gray frames identify the unmeasured output ports (4, 6, and 8).
		\subref{fig:intro-input} Random input sequence $u_n$ encoded in optical power.
		\subref{fig:intro-responses} Representative MRR network responses to the common programmed input.
		The green trace is the port-1 out-of-resonance reference; the orange and blue traces are port-1 responses at two resonant laser frequencies, and the pink trace is a port-5 response.
		\subref{fig:intro-metrics} Overview of selected analyses, namely temporal spectral analysis, memory-capacity characterization, and NARMA reconstruction.
	}
	\label{fig:intro}
\end{figure}

The chip we will be using contains both a single add-drop MRR as well as a fully passive $8\times8$ nominally equal MRR network, allowing us to compare the dynamics of a single resonator and of a network.
Both devices are fabricated on a silicon-on-insulator platform with silica waveguides of cross-section $\qty{450}{\nano\meter}\times\qty{220}{\nano\meter}$.
The single MRR and each MRR in the network share a racetrack geometry with a bend radius of \qty{7}{\micro\meter}.
Device fabrication and geometry are detailed in \cref{sec:methods-circuit}.
The single MRR is shown in \cref{fig:intro-ring}.
The single MRR is shown in \cref{fig:intro-ring}.
The MRR network on the other hand has one physical input and nine physical output ports, as shown in \cref{fig:intro-matrix}.
Fabrication-induced differences distribute the resonance frequencies across the MRR network \cite{bogaerts2011silicon,biasi2024photonic}, and no individual MRR is actively tuned after fabrication.

\subsection{Random Excitation of the Free-Carrier--Thermal Dynamics}\label{sec:input-output-encoding}

To show how the systems respond to a continuously changing input, we therefore encode a discrete random sequence $u_n$ in the optical input power, as illustrated in \cref{fig:intro-input}.
The symbol duration $T$ sets how long each random input level is applied.
We use $T=\qtylist{10;20;40}{\nano\second}$ to vary the modulation timescale relative to the free-carrier and thermal responses while maintaining temporal overlap between consecutive input-driven transients.
This overlap is necessary to probe fading memory because the random sequence includes input levels with zero optical power.
If such a level were held for much longer than the relevant relaxation times, the free-carrier--thermal perturbations and measured output would relax toward their unexcited states before the next symbol.
The zero-power interval would then act as a reset that erases the observable influence of preceding inputs rather than as one step in a continuous fading-memory sequence.
The selected symbol durations therefore limit this interruption while allowing us to test how the input-modulation timescale affects the response.
The resulting output traces provide the common basis for the spectral, memory, and task analyses.

The waveform generation, power calibration, output sampling, and input--output alignment are described in \cref{sec:methods-input,sec:methods-timing}.

\subsection{Wavelength- and Port-Diverse Observations}\label{sec:wavelength-port-acquisition}

The response of an MRR depends on the laser frequency and input power, and the output of each MRR network output port is a consequence of a different combination of resonances and propagation paths.
We therefore repeat the same programmed random input across laser-frequency--power operating points and measured output ports.
Because every acquisition uses the same input sequence, the resulting traces share a common timeline and can be compared at matched sequence positions.

The laser frequencies and output ports are measured sequentially and combined only in post-processing.
This procedure does not demonstrate simultaneous multi-wavelength or multiport acquisition.
Instead, it tests whether separately measured responses expose complementary information that is weak or absent in any one response.
Out-of-resonance traces provide approximately linear and memoryless references under the measured conditions.

The operating-point grids, reference frequencies, acquisition sequence, and feature-combination procedures are detailed in \cref{sec:methods-acquisition,sec:methods-mc-feature-combination}.

\subsection{From Observable Dynamics to Processing Properties}\label{sec:characterization-metrics}

The measured traces contain several aspects of the response that no single analysis can separate.
We therefore proceed from the physical signature that is easiest to observe to progressively more demanding tests of temporal processing.
Selected stages of this sequence are illustrated in \cref{fig:intro-metrics}.
First, temporal spectra test whether the output-oscillation component associated with constant-drive self-pulsing remains visible under random modulation.
Memory-capacity (MC) analysis then locates and quantifies linearly reconstructable delayed input information \cite{dambre2012information}.
We use the decay of the delay-resolved MC profile to estimate effective relaxation scales.
Information-processing capacity (IPC) generalizes MC by testing the reconstruction of polynomial functions of one delayed input.
Finally, NARMA-$N$ reconstruction tests whether the measured responses provide sufficient memory and nonlinear interactions among different input symbols to reproduce a target sequence.

\subsubsection{Persistence of the Self-Pulsing-Associated Component}\label{sec:principle-spectral}

We first compare constant and random excitation at nominally matched mean optical power.
A resolved output oscillation under constant excitation identifies self-pulsing, but random modulation adds a broadband input-driven background.
Thus, we examine the random-modulation spectra for a corresponding narrow frequency component.
We call such a randomly driven response self-pulsing-like rather than self-pulsing because its dynamics are also driven by the changing input.
Here, the self-pulsing (SP) frequency refers to the frequency of the measured output oscillation and is distinct from the optical laser frequency that defines the operating point.
We then compare random-excitation spectra acquired at the same laser frequency and mean optical input power but with different symbol durations.
This comparison tests whether the self-pulsing-associated spectral component persists when the input-modulation timescale changes.

The spectral calculation and empirical response classification are described in \cref{sec:methods-spectral-analysis}.

\subsubsection{Fading Memory and Effective Relaxation Scales}\label{sec:principle-mc}

The spectral comparison reveals persistent temporal components, but it does not show whether the current output retains information about previous inputs.
Memory-capacity analysis measures how accurately a linear readout of the current response reconstructs earlier inputs \cite{dambre2012information}.
At each physical time lag $\tau_k$, the coefficient $\mathrm{MC}_k$ quantifies the reconstruction of the corresponding delayed input, and total MC aggregates the delayed profile after removal of the fitted instantaneous contribution.
The function $\mathrm{MC}_k$ versus $\tau_k$ is the delay-resolved memory profile and shows how the reconstructable input information decays with physical delay.

We first map MC for individual responses to locate the operating conditions with delayed information.
We then combine responses measured at different laser frequencies or output ports to test whether they expose complementary input history.
Finally, we fit the fast and slow decay regions of this profile to obtain effective input--output relaxation scales for comparison with free-carrier--thermal dynamics.

The MC calculation, acquisition-broadening normalization, response combination, decay fitting, and simulation validation are described in \cref{sec:methods-mc-analysis,sec:methods-mc-feature-combination,sec:methods-mc-decay-fits,sec:methods-mc-simulation}.

\subsubsection{Nonlinear Transformations of Delayed Inputs}\label{sec:principle-ipc}

MC identifies linearly reconstructable input history, but it does not determine whether nonlinear transformations of that history are also accessible.
Instead, information-processing capacity (IPC) replaces the delayed-input target with polynomial functions of one delayed input \cite{dambre2012information}.
The first-order capacity is the linear MC contribution, while higher orders test nonlinear transformations of increasing polynomial degree.
This comparison reveals whether the measured representation is dominated by linear memory or also provides reconstructable nonlinear transformations.
Because each target depends on only a delayed input, this IPC analysis does not test nonlinear interactions between inputs at different delays, such as $u_{n-k_1}u_{n-k_2}$ with $k_1\neq k_2$.

The IPC targets, normalization, and aggregation are described in \cref{sec:methods-ipc-analysis}.

\subsubsection{NARMA-\texorpdfstring{$N$}{N} Reconstruction}\label{sec:principle-narma}

\begin{figure}[htp]
	\includegraphics[width=\linewidth]{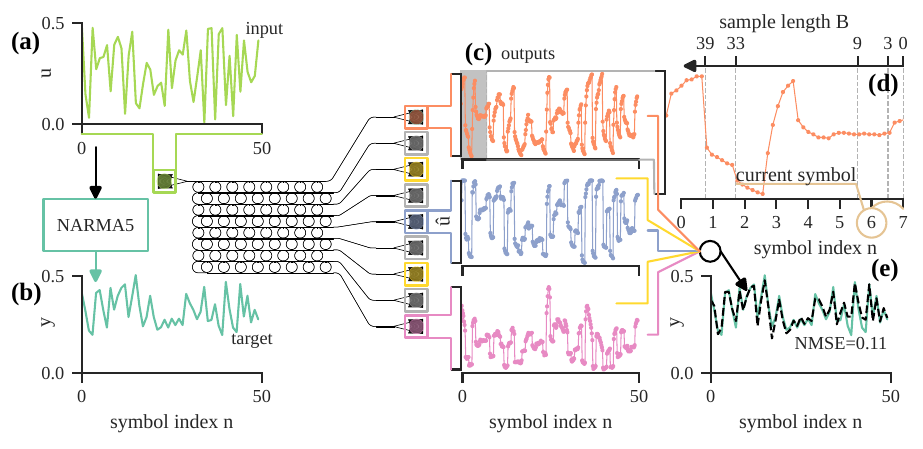}

	\begin{subcaptiongroup}
		\phantomcaption\label{fig:narma-explanation-input}
		\phantomcaption\label{fig:narma-explanation-target}
		\phantomcaption\label{fig:narma-explanation-port-responses}
		\phantomcaption\label{fig:narma-explanation-buffers}
		\phantomcaption\label{fig:narma-explanation-reconstruction}
	\end{subcaptiongroup}

	\caption{%
		\textbf{NARMA target generation, photonic-response sample-window construction, and linear reconstruction.}
		\subref{fig:narma-explanation-input} Random input sequence $u_n$ supplied to the NARMA5 recurrence and the MRR network.
		\subref{fig:narma-explanation-target} NARMA5 target $y_n$ calculated from the input sequence.
		\subref{fig:narma-explanation-port-responses} Responses measured at MRR network ports 1, 5, and 9, shown in orange, blue, and pink, respectively; dots mark the acquired output samples.
		\subref{fig:narma-explanation-buffers} Enlarged port-1 response illustrating windows of length $B$. These windows are right-aligned, i.e. each window ends at the same causal readout time where $B=0$. E.g., a window with sample length $B=3$ contains the first three output samples that are assigned to the current symbol, and the longer windows extend backward into preceding-symbol intervals.
		\subref{fig:narma-explanation-reconstruction} NARMA5 target (aqua) and linear-readout reconstruction (dashed black) formed from the measured responses at ports $\{1,3,5,7,9\}$.
	}
	\label{fig:narma-explanation}
\end{figure}

Finally, we use NARMA-$N$ reconstruction to test whether the measured responses combine fading memory with nonlinear interactions among different input symbols.
Specifically, the target sequence depends on delayed inputs, products of input symbols, and previous target values \cite{atiya2000new,wringe2025reservoir}.
\Cref{fig:narma-explanation-input,fig:narma-explanation-target} show the input sequence and the corresponding NARMA target.
NARMA5 is the primary benchmark, and the other evaluated orders vary the required temporal history and recurrence order.

Here, a symbol is one element $u_n$ of the programmed input sequence, while a sample is one acquired value of a measured output response, which do not necessarily have to be the same, see \cref{fig:narma-explanation-port-responses,fig:narma-explanation-buffers}.
Unlike the MC analysis, the NARMA analysis retains the original symbol-indexed input sequence instead of resampling it at the shorter output-sample times.
This avoids interpolation-related mixing between neighboring symbols and the associated broadening discussed in \cref{sec:methods-mc-analysis}.
Retaining the symbol-indexed sequence also allows the samples associated with each symbol to test whether the induced output dynamics, including a self-pulsing-associated component, contribute to NARMA reconstruction.
Retaining the sample-indexed sequence also allows the samples associated with each symbol to test whether the induced output dynamics, including a self-pulsing-associated component, contribute to NARMA reconstruction.
The sample length $B$ is the number of consecutive output samples supplied to the readout.
The maximum symbol depth $D$ identifies the most distant preceding symbol associated with these samples, and $D=0$ denotes the current-symbol condition.
\Cref{fig:narma-explanation-port-responses,fig:narma-explanation-buffers} show the measured port responses and the construction of the sample windows.
We select each sample length to realize a specified maximum symbol depth.
For the primary $T=\qty{40}{\nano\second}$ benchmark, $B=3$ corresponds to $D=0$ and uses the first three samples assigned to each symbol.
For example, the sample-window duration for a single symbol corresponds to $B=3$, and is $B\Delta t=\qty{19.2}{\nano\second}<T/2$, so these samples are unlikely to include AWG interpolation toward the next symbol.
Increasing the sample length $B$ extends the sample window $B\Delta t$ into preceding-symbol intervals, increases $D$, and therefore defines a hybrid photonic--digital processor.
\Cref{fig:narma-explanation-reconstruction} compares the NARMA target with the reconstruction obtained from the measured responses.

The recurrence, sample-window alignment, feature construction, digital reference, data partitioning, and performance evaluation are described in \cref{sec:methods-narma-analysis}.

\section{Results and Discussion}\label{sec:results}

Next, we will discuss the results of our investigations. \Cref{sec:results-spectral} shows that a dominant spectral component associated with constant-drive self-pulsing remains observable under random modulation and persists over the investigated symbol durations.
\Cref{sec:results-memory} shows that the largest MC values occur near the constant-drive self-pulsing boundaries, that the measured-response combination increases information accessible to the readout, and that MC decay provides effective relaxation-scale estimates.
\Cref{sec:results-ipc} shows that the MRR network representation contains predominantly linear delayed input information and a smaller quadratic single-delay contribution.
\Cref{sec:results-narma} shows that current-symbol photonic responses contain information about preceding inputs, but the measured MRR network does not provide all the temporal history required by the evaluated NARMA tasks under the investigated conditions.
Increasing the sample length to include samples associated with preceding symbols supplies this additional history digitally and improves reconstruction.

\subsection{Spectral Dynamics under Constant and Random Excitation}\label{sec:results-spectral}

\subsubsection{Self-Pulsing of the Single MRR}\label{sec:results-spectral-ring}

Here, we show that the randomly modulated response of the single MRR retains a dominant temporal spectral component near the self-pulsing frequency measured under constant excitation.

\begin{figure}[htp]
	\includegraphics[width=\linewidth]{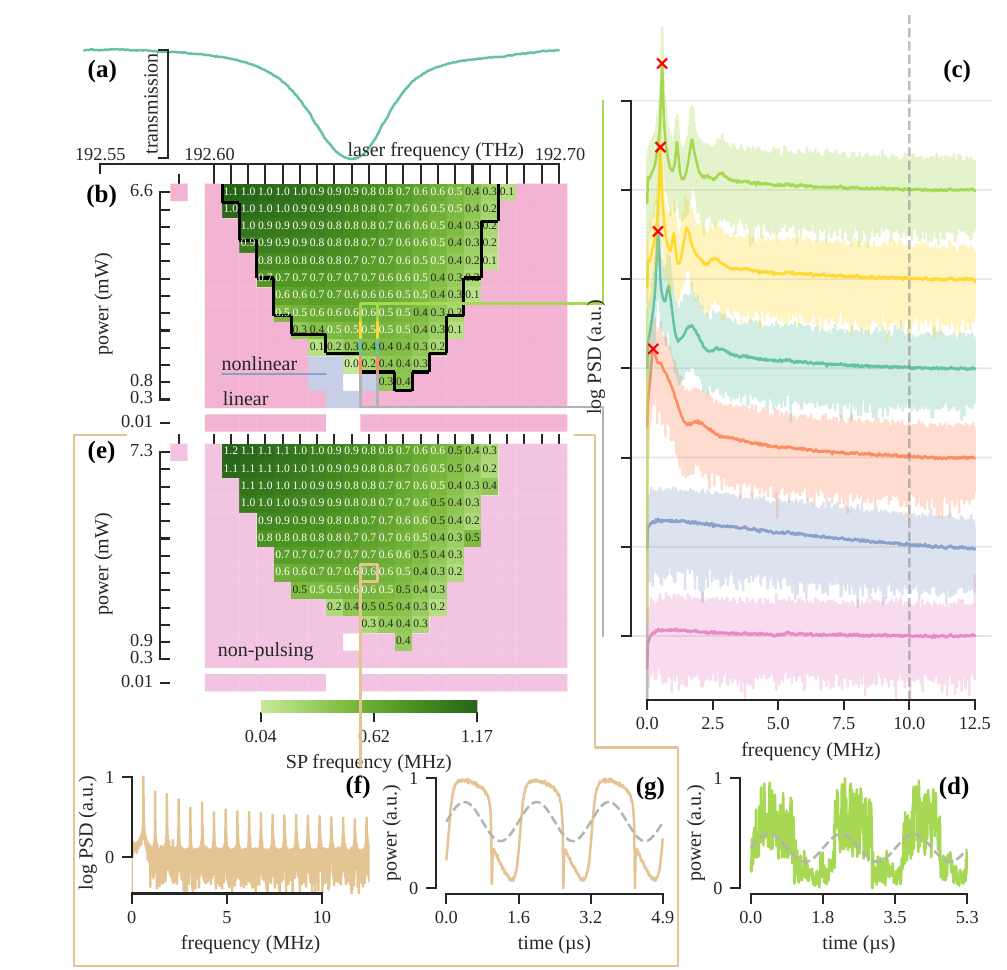}

	\begin{subcaptiongroup}
		\phantomcaption\label{fig:fft-ring-spectrum}
		\phantomcaption\label{fig:fft-ring-random-map}
		\phantomcaption\label{fig:fft-ring-random-psd}
		\phantomcaption\label{fig:fft-ring-random-signal}
		\phantomcaption\label{fig:fft-ring-constant-map}
		\phantomcaption\label{fig:fft-ring-constant-psd}
		\phantomcaption\label{fig:fft-ring-constant-signal}
	\end{subcaptiongroup}

	\caption{%
		\textbf{Measured through-port spectral responses of the single MRR under random and nominally mean-power-matched constant excitation.}
		\subref{fig:fft-ring-spectrum} Low-power optical spectrum; \qty{192.55}{\tera\hertz} is the out-of-resonance reference frequency.
		\subref{fig:fft-ring-random-map} Empirical response classes for random excitation with $T=\qty{10}{\nano\second}$ as a function of laser frequency and mean optical power.
		Pink, light blue, and green cells denote approximately linear input-like responses, nonlinear responses without a resolved self-pulsing-like signature, and self-pulsing-like responses, respectively.
		The green shade and number in each green cell give the dominant output-oscillation frequency; white cells denote acquisitions that could not be parsed, and the black contour is the constant-drive self-pulsing boundary mapped onto the measured random-excitation power axis.
		\subref{fig:fft-ring-random-psd} Vertically offset log-PSD spectra at a laser frequency of \qty{192.645}{\tera\hertz} for increasing mean optical powers; red crosses mark maxima of the smoothed log PSD, and the gray dashed line at \qty{10}{\mega\hertz} marks the lower boundary of the \qtyrange{10}{12.5}{\mega\hertz} normalization interval.
		\subref{fig:fft-ring-random-signal} Temporal response corresponding to the highest-power spectrum in \subref{fig:fft-ring-random-psd}, at a mean optical power of \qty{2.9}{\milli\watt}; the dashed sinusoid uses the displayed dominant output-oscillation frequency of approximately \qty{0.57}{\mega\hertz}.
		\subref{fig:fft-ring-constant-map} Constant-excitation classification over the same laser-frequency and nominal power settings; green and pink cells denote self-pulsing and nonperiodic responses, respectively, and the green shade and number give the self-pulsing frequency.
		\subref{fig:fft-ring-constant-psd}--\subref{fig:fft-ring-constant-signal} Log PSD and temporal response at \qty{192.645}{\tera\hertz} and \qty{3.1}{\milli\watt}; the dashed sinusoid uses the displayed self-pulsing frequency of \qty{0.61}{\mega\hertz}.
	}
	\label{fig:fft-ring}
\end{figure}

To test this persistence, we acquire random- and constant-excitation responses on similar laser-frequency--power grids around the single-MRR resonance.
The low-power through-port spectrum in \cref{fig:fft-ring-spectrum} identifies this resonance and provides the spectral context for the sweep.

\Cref{fig:fft-ring-random-map,fig:fft-ring-constant-map} show a strong qualitative overlap between the random- and constant-excitation maps.
However, the measured random-excitation powers are approximately \qty{90}{\percent} of the corresponding constant-excitation powers, so we account for this mismatch before comparing the output-oscillation frequencies.
Across the $21\times14=294$ grid cells, excluding the out-of-resonance reference, 130 are self-pulsing under constant excitation.
Of the 138 random-excitation self-pulsing-like states, 129, or \qty{93.5}{\percent}, have a self-pulsing constant state at one or both adjacent constant-power bounds.
For the 112 states bounded by two constant-drive self-pulsing states, we assume that the self-pulsing frequency varies linearly with power between the two bounds.
Across these 112 states, the mean absolute difference between the random-modulation frequency and the power-interpolated constant-drive frequency is \qty{0.0227}{\mega\hertz}, and the mean absolute relative difference is \qty{4.18}{\percent}.
Most of these states lie within the bulk of the constant-drive self-pulsing region: for these 96 states, the mean absolute difference is \qty{0.0197}{\mega\hertz} and the mean absolute relative difference is \qty{3.07}{\percent}.
For the 16 states adjacent to the self-pulsing power boundary, the corresponding differences increase to \qty{0.0406}{\mega\hertz} and \qty{10.8}{\percent}.
Thus, the frequency agreement weakens near the self-pulsing boundary but remains close for most states, indicating that the temporal component associated with constant-drive self-pulsing remains observable under random modulation.

We classify the random-excitation traces into three empirical response classes.
The classification uses the area under the normalized logarithm of the power spectral density (log PSD) and the exponential-envelope fit error.
The thresholds are calibrated from the single-MRR metric distributions, as detailed in \cref{sec:methods-spectral-analysis}.
In \cref{fig:fft-ring-random-map}, green cells identify self-pulsing-like responses, and their numerical values report the dominant output-oscillation frequency in \unit{\mega\hertz}.
Pink cells identify empirical approximately linear input-like responses, and light blue cells identify empirical nonlinear responses without this resolved component.
The black contour marks the constant-drive self-pulsing boundary after matching its measured mean-power coordinates to the random-excitation power axis.
In \cref{fig:fft-ring-constant-map}, green cells identify self-pulsing responses and pink cells identify nonperiodic responses.
White cells identify acquisitions that could not be parsed reliably.

To connect the map-level result to individual measured traces, we examine representative random- and constant-excitation responses.
\Cref{fig:fft-ring-random-psd,fig:fft-ring-random-signal} show the retained spectral component and its oscillatory envelope under random modulation.
The approximately flat out-of-resonance random-input PSD is qualitatively similar to the lowest-power spectrum in \cref{fig:fft-ring-random-psd}, so we do not show it separately.
\Cref{fig:fft-ring-constant-psd,fig:fft-ring-constant-signal} show the corresponding spectral component and periodic response under constant excitation.
In the random temporal response, the retained component forms a relatively slow oscillatory envelope produced by the coupled free-carrier--thermal dynamics.
Faster fluctuations driven by the random input are superimposed on this envelope and contribute mainly to the broadband spectral background.
The narrow component therefore tracks the free-carrier--thermal oscillation, but random modulation makes the higher harmonics less resolved.

The maps also reveal nonlinear responses without a resolved self-pulsing-like component near the constant-drive self-pulsing island.
Their output differs from the approximately linear input-like response, but their spectra contain no narrow persistent component that meets the self-pulsing-like criterion.
The constant-excitation map does not identify this input-dependent nonlinear behavior because it classifies only periodic and nonperiodic responses.
This limitation motivates the fading-memory analysis, which tests whether these responses retain information about previous inputs.

\subsubsection{Spectral Dynamics of the MRR Network}\label{sec:results-spectral-matrix}

The MRR network exhibits observable spectral behavior similar to the single MRR, and its resonant temporal components persist over the investigated symbol durations.
We therefore present port 1 as a representative example; qualitatively similar behavior is observed at the other measured output ports.

\begin{figure}[htp]
	\includegraphics[width=\linewidth]{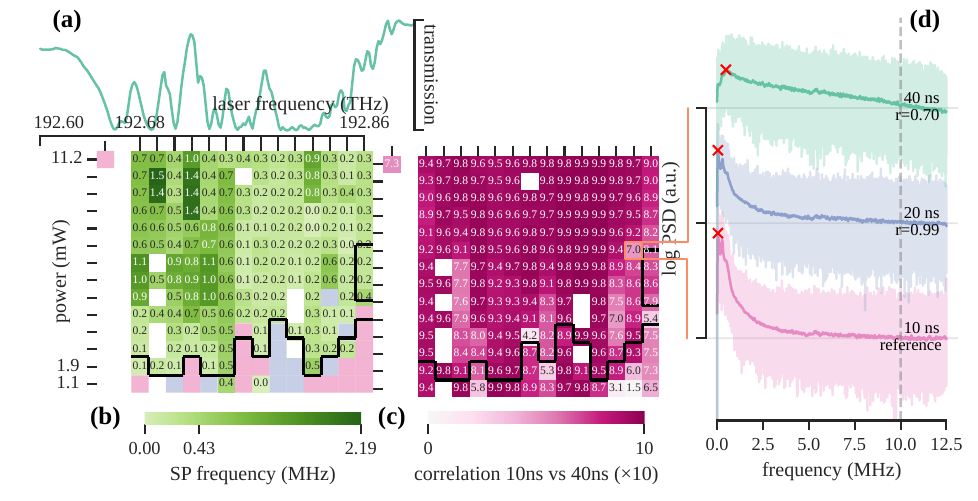}

	\begin{subcaptiongroup}
		\phantomcaption\label{fig:fft-matrix-spectrum}
		\phantomcaption\label{fig:fft-matrix-random-map}
		\phantomcaption\label{fig:fft-matrix-correlation}
		\phantomcaption\label{fig:fft-matrix-timestep-psd}
	\end{subcaptiongroup}

	\caption{%
		\textbf{Measured spectral responses at port 1 of the MRR network under random excitation.}
		\subref{fig:fft-matrix-spectrum} Low-power optical spectrum; \qty{192.60}{\tera\hertz} is the out-of-resonance reference frequency.
		\subref{fig:fft-matrix-random-map} Empirical response classes for random excitation with $T=\qty{10}{\nano\second}$ as a function of laser frequency and mean optical power, using the thresholds calibrated on the single MRR.
		Colors follow \cref{fig:fft-ring-random-map}.
		\subref{fig:fft-matrix-correlation} Correlation between smoothed spectra measured at matched operating points for $T=\qty{10}{\nano\second}$ and $T=\qty{40}{\nano\second}$; cell annotations report $10r$, where $r$ is the Pearson correlation coefficient.
		The outlined example at \qty{192.846}{\tera\hertz} and \qty{7.3}{\milli\watt} has $r\approx\num{0.99}$ for \qty{10}{\nano\second} versus \qty{20}{\nano\second} and $r\approx\num{0.70}$ for \qty{10}{\nano\second} versus \qty{40}{\nano\second}.
		\subref{fig:fft-matrix-timestep-psd} Vertically offset log-PSD spectra at the outlined operating point for $T=\qtylist{10;20;40}{\nano\second}$, shown from bottom to top.
		Red crosses mark FFT maxima at approximately \qtylist{0.048;0.045;0.477}{\mega\hertz}, respectively.
	}
	\label{fig:fft-matrix}
\end{figure}

The low-power transmission spectrum in \cref{fig:fft-matrix-spectrum} provides the spectral content for the port-1 measurements.
Its multiple features are consistent with contributions from several resonances and propagation paths that reach this output.

We test whether the three empirical response classes identified for the single MRR are also present in the MRR network responses.
To enable direct comparison, we classify the responses measured at port 1 and the other output ports using thresholds calibrated on the single-MRR data.
We visually inspected the classified traces and verified that the labels identify the intended spectral signatures.
\Cref{fig:fft-matrix-random-map} shows the representative port-1 classification using the color convention of \cref{fig:fft-ring-random-map}.
The occurrence of the same response classes provides evidence that the MRR network and single MRR exhibit similar observable spectral regimes.
The black contour is derived from constant-drive MRR network traces.

To determine whether the resonant temporal components persist when the input-modulation timescale changes, we calculate the signed Pearson correlation $r$ between smoothed spectra measured at matched laser frequency and mean optical power.
We first compare $T=\qty{10}{\nano\second}$ with $T=\qty{20}{\nano\second}$ and then increase the difference in modulation timescale by comparing $T=\qty{10}{\nano\second}$ with $T=\qty{40}{\nano\second}$.
Both comparisons contain common random-modulation states, including 156 states within the constant-drive self-pulsing subset.
For the smaller change in symbol duration, the mean correlation within this subset is $r=\num{0.982}$, and 154 of 156 states, or \qty{98.7}{\percent}, satisfy the high-correlation criterion $r>\num{0.90}$.
When the comparison is extended to $T=\qty{40}{\nano\second}$, the mean correlation decreases to $r=\num{0.930}$, but 125 of 156 states, or \qty{80.1}{\percent}, remain above the same threshold.

Most constant-drive self-pulsing states therefore retain a closely related spectral shape across the tested symbol durations, although the agreement decreases as the modulation timescales increases.
Increasing the symbol duration prolongs the intervals during which the randomly modulated input power remains below its mean value.
Constant excitation at the same mean power contains no such low-power excursions.
These longer low-power intervals may suppress the self-pulsing-associated dynamics and explain the disappearance of the resolved spectral component at some operating points.

\Cref{fig:fft-matrix-correlation} maps the \qty{10}{\nano\second}-to-\qty{40}{\nano\second} comparison and shows where this persistence occurs in the laser-frequency--power plane.
At the out-of-resonance reference frequency, the correlation is low because the approximately linear response follows the random input, whose spectral background changes with symbol duration.

The aggregate correlation analysis shows that the spectral component usually persists but disappears at some operating points, so we inspect one such case directly.
\Cref{fig:fft-matrix-timestep-psd} shows the log-PSD spectra at the outlined operating point for $T=\qtylist{10;20;40}{\nano\second}$.
The \qty{10}{\nano\second} and \qty{20}{\nano\second} spectra are highly correlated and retain low-frequency maxima at approximately \qtylist{0.048;0.045}{\mega\hertz}.
When the symbol duration increases to \qty{40}{\nano\second}, this low-frequency peak disappears, the retained FFT maximum moves to approximately \qty{0.477}{\mega\hertz}, and the correlation with the \qty{10}{\nano\second} spectrum decreases to $r\approx\num{0.70}$.
This example shows directly how the self-pulsing-associated spectral component can be lost at the operating points whose correlation decreases for the larger symbol duration.

\subsection{Fading Memory and Effective Relaxation-Scale Estimation}\label{sec:results-memory}

All MC results in this subsection use responses acquired with $T=\qty{10}{\nano\second}$.
The spectral analysis we performed so far identifies self-pulsing-like and nonoscillatory nonlinear responses, but it does not reveal how long these responses retain information about the random input.
We therefore use the delay-resolved memory capacity $\mathrm{MC}_k$ to map fading memory in the single MRR and in the MRR network.
After establishing these memory landscapes, we test whether combining responses measured at different laser frequencies and output ports exposes additional delayed information, and we use the MC decay to estimate effective relaxation scales.

\subsubsection{Memory Landscape of the Single MRR}\label{sec:results-memory-ring}

\begin{figure}[htp]
	\includegraphics[width=\linewidth]{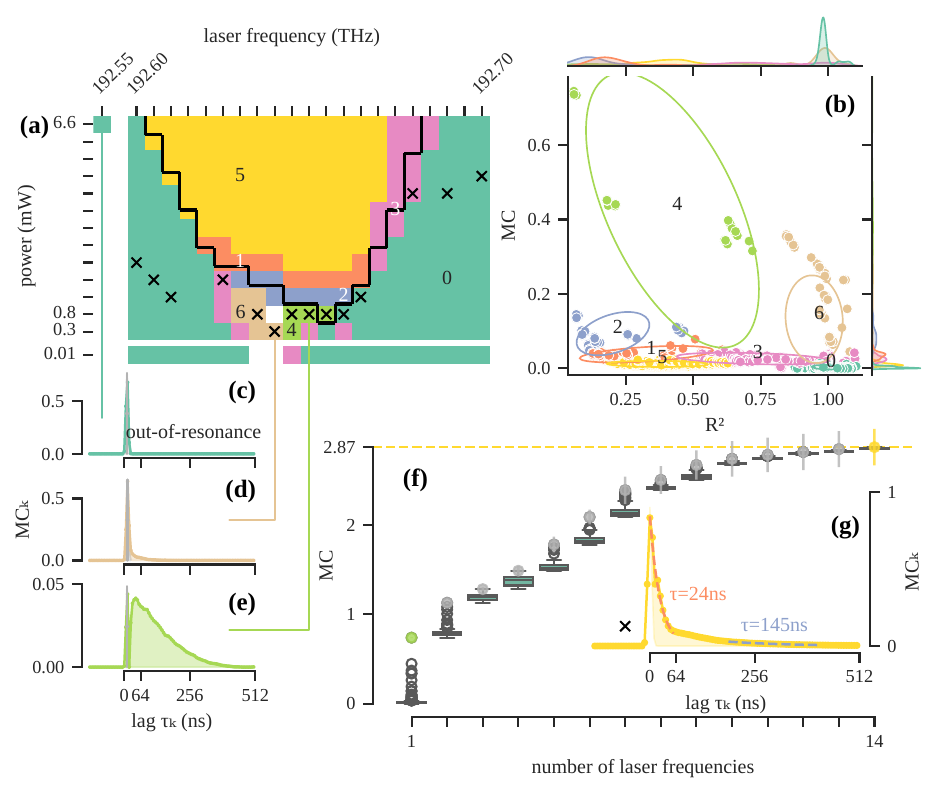}

	\begin{subcaptiongroup}
		\phantomcaption\label{fig:mc-ring-map}
		\phantomcaption\label{fig:mc-ring-gmm}
		\phantomcaption\label{fig:mc-ring-example-out-of-resonance}
		\phantomcaption\label{fig:mc-ring-example-fast}
		\phantomcaption\label{fig:mc-ring-example-slow}
		\phantomcaption\label{fig:mc-ring-frequency-combination}
		\phantomcaption\label{fig:mc-ring-maximum-profile}
	\end{subcaptiongroup}

	\caption{%
		\textbf{Memory capacity space from measured through-port responses of the single MRR and from laser-frequency-diverse representations at $T=\qty{10}{\nano\second}$.}
		\subref{fig:mc-ring-map} Empirical Gaussian-mixture-model (GMM) groups over laser frequencies and mean optical powers.
		Colors and integers identify the seven GMM groups, white cells denote excluded acquisitions, the black contour is the constant-drive self-pulsing boundary, black crosses mark operating points retained by the greedy combination used in \subref{fig:mc-ring-frequency-combination}.
		\subref{fig:mc-ring-gmm} Operating points in instantaneous-contribution $R^2$ -- memory-capacity MC space.
        Colored curves along the top and right margins show the marginal density distributions of $R^2$ and MC, respectively, for each GMM group.
		\subref{fig:mc-ring-example-out-of-resonance}--\subref{fig:mc-ring-example-fast}--\subref{fig:mc-ring-example-slow} Delay-resolved profiles for the out-of-resonance reference and representative fast-decay and slow-decay responses, respectively.
        \subref{fig:mc-ring-frequency-combination} Greedy combination of sequentially measured laser-frequency responses.
		The procedure first selects the laser-frequency--power response with the largest mean MC.
		It then tests the addition of each response at the remaining laser frequencies and retains the combination with the largest mean MC, repeating this process at each step.
		Candidate circles show the mean MC of each tested combination across eight independent input series, and box plots summarize the distribution of these means.
		Highlighted markers identify the selected combinations, and their error bars give the standard deviation across the eight series.
		The yellow dashed line marks the final selected mean.
		\subref{fig:mc-ring-maximum-profile} Delay-resolved profile of the selected 14-frequency representation; the orange and blue dashed curves are exponential fits to the displayed fast and slow decay regions.
	}
	\label{fig:mc-ring}
\end{figure}

We first map where the single MRR provides current and delayed information about the random input.
The normalized instantaneous contribution $R^2$ (refer to description in \cref{sec:methods-mc-analysis}) quantifies linear reconstruction of the current input, and MC quantifies linear reconstruction of previous inputs.
A high $R^2$ on the other hand indicates accurate linear reconstruction of the current input, while a low $R^2$ indicates poor current-input reconstruction.
A high MC indicates that preceding inputs remain linearly reconstructable, while MC near zero indicates little accessible delayed information.
\Cref{fig:mc-ring-map,fig:mc-ring-gmm} reveal three broad reconstruction regimes across the sampled operating grid.
Outside the constant-drive self-pulsing boundary, many operating points have high $R^2$ and MC near zero.
These responses provide an approximately linear and memoryless transformation, consistent with the input-like responses identified by the spectral analysis.
Inside the region enclosed by the constant-drive self-pulsing boundary, both $R^2$ and MC are generally low.
The resolved oscillatory dynamics in this region are therefore associated with poor linear reconstruction of both the current and previous inputs under the investigated conditions.
Near the boundary but outside the self-pulsing region, several operating points have appreciable MC values.
Compared with the other regions, these operating points therefore better support linear reconstruction of previous inputs.

The reconstruction properties vary within these broad regimes, so we use empirical groups to summarize the memory landscape.
\Cref{fig:mc-ring-gmm} shows seven groups identified using a Gaussian mixture model (GMM), including responses dominated by instantaneous input reconstruction, responses with little reconstructable information, and responses with appreciable delayed memory.
These groups describe similar observable reconstruction properties.
The GMM construction and normalization are detailed in \cref{sec:methods-mc-analysis}.

However, scalar total MC does not show how the reconstructable information decays with physical delay.
Instead, each $\mathrm{MC}_k$ coefficient measures the linear component of memory accessible from the detected response at delay $k$.
The raw coefficient $\mathrm{MC}_{k,\mathrm{raw}}=1$ corresponds to perfect linear reconstruction of the delayed input $u_{n-k}$.
After normalization by the out-of-resonance instantaneous contribution, $\mathrm{MC}_k=0$ means that the response contains no linearly reconstructable trace of that input.
The latter can occur in two conceptually different cases: the response may encode the delayed input only through a nonlinear transformation that is inaccessible to the linear readout, or the detector may receive no input-dependent output.
For example, at zero launched optical power, no input-dependent optical signal reaches the detector and every $\mathrm{MC}_k$ must be zero.
Therefore, a low $\mathrm{MC}_k$ does not by itself distinguish a nonlinear encoding from the absence of an observable response.
For example, the out-of-resonance profile in \cref{fig:mc-ring-example-out-of-resonance} has $\mathrm{MC}_k \approx 0$ for every nonzero delay index $k$, as expected for a linear but memoryless response.
To illustrate distinct memory timescales, \cref{fig:mc-ring-example-fast,fig:mc-ring-example-slow} show two representative delay-resolved profiles.
We selected these examples because the first is dominated by a rapidly decaying component, while the second retains a weaker contribution over a longer physical delay.
Scalar total MC subtracts the fitted instantaneous contribution from the sum of the normalized delay-resolved profile and therefore quantifies the aggregate delayed information accessible to the linear readout.
However, it does not show at which delays this information is available or how strongly each delay contributes.

Up to this point, MC is evaluated from one response trace acquired at a single laser-frequency--power operating point.
We now combine traces measured at different laser frequencies and treat them as multiple representations of the single MRR in one enlarged feature vector.
We evaluate each candidate combination separately for eight independent random input series and rank the combinations by their mean MC.
The greedy selection starts with the laser-frequency--power operating point with the largest mean MC and then adds the response that gives the largest mean MC when combined with those already selected.
The procedure retains at most one operating power per laser frequency and is detailed in \cref{sec:methods-mc-greedy-frequency}.
The benefit of this expanded representation appears in \cref{fig:mc-ring-frequency-combination}.
MC increases from $0$ for the out-of-resonance, approximately linear reference to $\mathrm{MC}=\num{0.738(4)}$ for the selected single response and to $\mathrm{MC}=\num{2.9(2)}$ for the final 14-frequency feature set.
Most selected frequency--power operating points, marked by crosses in \cref{fig:mc-ring-map}, lie near the constant-drive self-pulsing boundary.
The improvement therefore comes mainly from combining edge responses that contain appreciable delayed memory.
As the selection proceeds, the MC increase gradually saturates because the available responses with substantial delayed information have already been included.
The remaining candidates are approximately linear and memoryless, so adding them provides little additional delayed input information.
The responses are acquired separately and combined during analysis, so this result describes laser-frequency-diverse post-processing rather than simultaneous WDM operation.
The available instrumentation does not support simultaneous multi-wavelength acquisition, so we cannot determine whether co-propagating wavelength channels would interact through the nonlinear MRR network dynamics or reproduce the same MC under simultaneous WDM operation.
Nevertheless, the sequential combination provides a qualitative approximation of the representational diversity that parallel wavelength channels could offer, provided that inter-channel nonlinear interactions do not substantially alter the individual responses.

Finally, \cref{fig:mc-ring-maximum-profile} shows the delay-resolved profile of the selected 14-frequency feature set.
The Gaussian fit separates the finite-width instantaneous contribution from the delayed profile.
Exponential fits over two selected delayed intervals give effective fast and slow MC-decay constants.

\subsubsection{MRR Network Memory and Measured-Response Diversity}\label{sec:results-memory-matrix}

\begin{figure}[htp]
	\includegraphics[width=\linewidth]{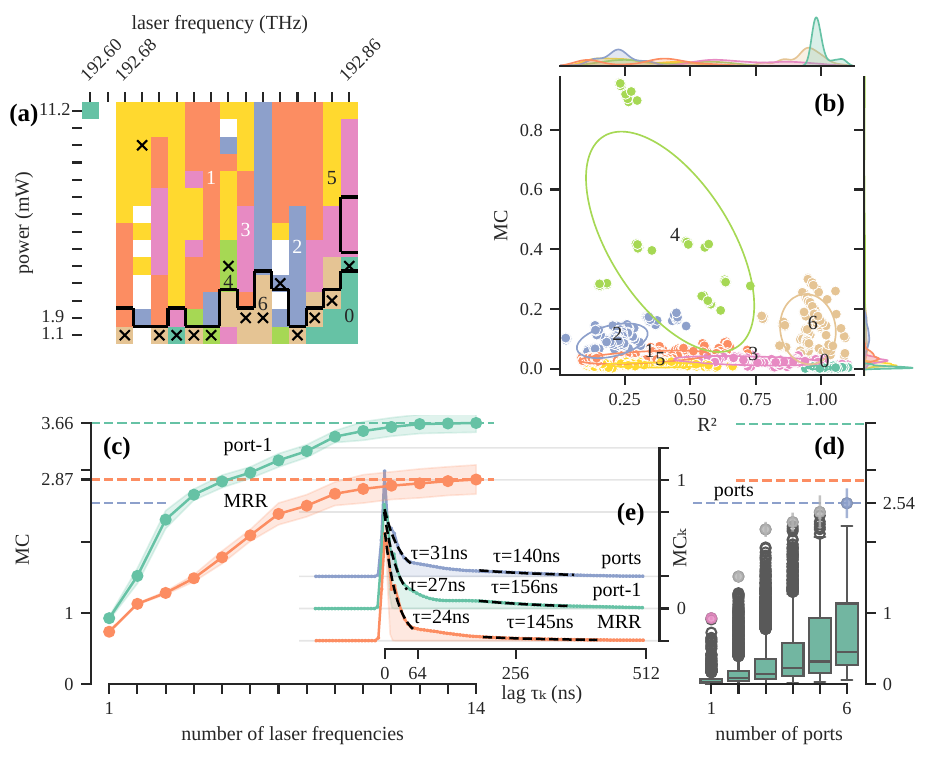}

	\begin{subcaptiongroup}
		\phantomcaption\label{fig:mc-matrix-map}
		\phantomcaption\label{fig:mc-matrix-gmm}
		\phantomcaption\label{fig:mc-matrix-frequency-comparison}
		\phantomcaption\label{fig:mc-matrix-port-combination}
		\phantomcaption\label{fig:mc-matrix-maximum-profiles}
	\end{subcaptiongroup}

	\caption{%
		\textbf{Memory capacity calculated from measured MRR network responses at $T=\qty{10}{\nano\second}$ and from laser-frequency- and output-port-diverse representations assembled in post-processing.}
		\subref{fig:mc-matrix-map} Empirical GMM groups at port 1 over laser frequencies and mean optical powers.
		Colors and integers identify the seven groups, white cells denote excluded acquisitions, the black contour is obtained from MRR network constant-excitation traces, black crosses mark selected operating points.
		\subref{fig:mc-matrix-gmm} Port-1 operating points in $R^2$--MC space.
        Colored curves along the top and right margins show the marginal density distributions of $R^2$ and MC, respectively, for each GMM group.
 		\subref{fig:mc-matrix-frequency-comparison} Greedy combinations of sequentially measured laser-frequency responses from the single-MRR through port (orange) and MRR-network port 1 (aqua).
		Circles show the largest mean MC selected at each step, averaged over eight independent input series.
		Shaded regions and vertical bars show the corresponding standard deviations across these series, and horizontal dashed lines mark the final selected means.
		\subref{fig:mc-matrix-port-combination} Combinations of the measured ports 1, 2, 3, 5, 7, and 9.
		All port subsets are tested separately at each laser-frequency--power operating point.
		For each port count, gray circles and box plots summarize the MC means across the tested port subsets and operating points, with each mean calculated over eight independent input series.
		Highlighted markers identify the largest mean MC for each port count, and their error bars show the corresponding standard deviations across the eight series.
		\subref{fig:mc-matrix-maximum-profiles} Delay-resolved profiles of the selected single-MRR, MRR-network-port-1, and six-port representations, shown in orange, aqua, and blue, respectively; black dashed curves are exponential fits to the displayed fast and slow regions.
	}
	\label{fig:mc-matrix}
\end{figure}

We now extend the memory analysis from the single MRR to the MRR network, first at port 1 and then across laser frequencies and output ports.
\Cref{fig:mc-matrix-map} shows that the largest MC values at port 1 again occur near the constant-drive self-pulsing boundary.
Thus, the port-1 memory landscape reproduces the main pattern observed for the single MRR.
The boundary is derived from constant-drive MRR network traces using thresholds calibrated on the single MRR.
We fit the GMM jointly to the single-MRR and port-1 MRR-network datasets.
\Cref{fig:mc-matrix-gmm} reports the resulting group assignments for the port-1 MRR-network operating points in $R^2$--MC space.
Therefore, operating points assigned the same group in the two datasets have comparable $R^2$ and MC values, which allows their current- and delayed-input reconstruction properties to be compared directly.

We apply the same greedy laser-frequency-combination procedure used for the single MRR to the responses measured at MRR network port 1.
\Cref{fig:mc-matrix-frequency-comparison} shows that the port-1 feature set increases from $\mathrm{MC}=\num{0.92(3)}$ for one response to $\mathrm{MC}=\num{3.6(1)}$ for the final combination.
The final MRR network representation contains more delayed input information than the corresponding representation of the single MRR, even though the network sweep uses a coarser laser-frequency spacing of approximately \qty{13.8}{\giga\hertz} compared with \qty{5}{\giga\hertz} for the single MRR.
This difference is consistent with the 64 resonators and multiple propagation paths providing a richer set of measured dynamical responses at port 1.

The MRR network also provides a second form of response diversity because its output ports sample different resonances and propagation paths.
We therefore expand the representation across physical output ports at fixed laser frequency and mean input power in \cref{fig:mc-matrix-port-combination}.
The selected maximum increases from $\mathrm{MC}=\num{0.92(3)}$ for one port to $\mathrm{MC}=\num{2.5(2)}$ when all six measured ports are combined.
This increase shows that the output ports provide nonidentical measured representations of the network dynamics.
The six port responses are currently acquired sequentially and aligned to the same input sequence.
However, this is a good approximation of a parallel measurement because the passive MRR network produces all output-port signals simultaneously, and detection at one output does not modify the signals at the other outputs.
A simultaneous detector array would therefore provide an equivalent port feature vector and MC result, apart from experimental noise and drift.
The exhaustive port-combination procedure is detailed in \cref{sec:methods-mc-simple-ports}.

Finally, \cref{fig:mc-matrix-maximum-profiles} compares the delay-resolved MC profiles of the selected single-MRR, port-1 MRR-network, and six-port MRR-network representations.
Combining responses measured at different laser frequencies accesses the MRR dynamics at different resonance detunings.
Combining output-port responses samples different optical paths and contributions from MRRs with slightly different resonance frequencies, thereby accessing a distribution of detunings.
These combinations change the profile amplitudes and total MC available to the readout, but the representative profiles retain visually similar fast and slow decay regions.
This similarity motivates the comparison of their effective decay constants, which we will discuss next in \cref{sec:results-relaxation-times}.

\subsubsection{Effective Relaxation Scales from Memory Decay}\label{sec:results-relaxation-times}

Here, our objective is to determine whether the measured MC decay can provide a direct input--output estimate of the fast and slow system timescales associated with free-carrier and thermal relaxation.
We therefore extract effective relaxation scales from the MC decay.

The selected memory functions in \cref{fig:mc-ring-maximum-profile,fig:mc-matrix-maximum-profiles} illustrate two delayed regions that are fitted separately with exponential decays.
The fast region follows the broadened instantaneous contribution, and the slower tail extends over hundreds of nanoseconds.
The curves in these figures are representative examples of the fitting procedure rather than the quantitative repeated-series estimates.

The single-response profiles in \cref{fig:mc-ring-example-fast,fig:mc-ring-example-slow} show that different measured representations expose different portions of the delay-resolved memory.
However, we cannot determine whether an individual representation captures the complete memory response, because some information may be absent from or inaccessible in that representation.
We therefore combine responses measured at different laser frequencies or output ports to maximize the delayed information accessible to the readout before fitting the fast and slow decay regions.
We analyze three selected representations: 14 laser frequencies for the single MRR, 14 laser frequencies for MRR network port 1, and six MRR network ports at one laser-frequency--power operating point.
The procedure in \cref{sec:methods-mc-decay-fits} performs these fits independently for each of the eight random input series.
\Cref{tab:mc-experimental-decays} reports the mean and standard deviation of the resulting effective MC-decay constants.
The single MRR, port-1 MRR network, and six-port MRR network configurations give similar mean fast and slow constants despite their different delayed-response amplitudes and total MC values.
The fitted MC-decay constants describe the decay of reconstructable input information.

\begin{table}[htp]
	\centering
	\caption{Effective MC-decay constants obtained from the experimental responses at $T=\qty{10}{\nano\second}$.
		Values are the mean and standard deviation of fits performed independently for eight random input series.
	}
	\begin{tabular}{lccc}
		Configuration        & $\tau_{\mathrm{MC,fast}}$ & $\tau_{\mathrm{MC,slow}}$ & $\textrm{MC}$ \\
		\hline
		Single MRR           & \qty{25(2)}{\nano\second}  & \qty{152(2)}{\nano\second} & \num{2.9(2)} \\
		MRR network port 1   & \qty{27(2)}{\nano\second}  & \qty{153(3)}{\nano\second} & \num{3.6(1)} \\
		Six-port MRR network & \qty{26(1)}{\nano\second}  & \qty{127(3)}{\nano\second} & \num{2.5(2)} \\
	\end{tabular}
	\label{tab:mc-experimental-decays}
\end{table}

We next derive the relation between an effective MC-decay constant and the corresponding physical relaxation time.
To interpret each fitted MC-decay constant, consider a system response with amplitude $x(t)$ that decays exponentially with physical relaxation time $\tau$: $x(t)\propto\exp(-t/\tau)$.
We assume that the reconstruction correlation at physical delay $\tau_k$ is proportional to this amplitude.
Because $\mathrm{MC}_k$ is the square of the reconstruction correlation, it decays as $\mathrm{MC}_k\propto\exp(-2\tau_k/\tau)$.
Comparing this expression with the fitted decay $\mathrm{MC}_k\propto\exp(-\tau_k/\tau_{\mathrm{MC}})$ gives the following relation.
\begin{equation}
	\tau_{\mathrm{MC}}\simeq\frac{\tau}{2}.
\end{equation}
Doubling the mean fitted constants in \cref{tab:mc-experimental-decays} gives candidate system-level relaxation times under this approximation.
It should be stressed that this method does not provide separate estimates of the physical rise and decay times.
Instead, it provides an effective timescale associated with the decay of reconstructable input information in the system.

We test this interpretation with a simulated single MRR based on the free-carrier--thermal dynamical equations and parameter sets reported by \textcite{vanvaerenbergh2012cascadable} and \textcite{borghi2021the}.
Both parameter sets are evaluated in a single MRR configuration.
The simulated responses use the same MC calculation and automatic decay-fitting procedure as the measurements.
The simulated memory profiles for both models exhibit distinct fast and slow decay regions similar to those observed experimentally.
\Cref{tab:mc-simulation-decays} compares twice the fitted MC-decay constants with the carrier and thermal relaxation times specified in the models.
The values are approximately comparable, consistent with the relation $\tau \simeq 2\tau_{\mathrm{MC}}$ under the exponential-response approximation.
These simulation results are consistent with assigning the fast and slow experimental memory components primarily to free-carrier and thermal relaxation, respectively.
Together, the factor-of-two agreement in both single MRR models and the comparability of the inferred timescales with carrier and thermal relaxation scales reported in the literature provide a strong indication that the fast and slow experimental MC-decay components are associated mainly with free-carrier and thermal relaxation, respectively.

\begin{table}[htp]
	\centering
	\caption{Comparison of specified physical relaxation times $\tau$ and twice the fitted MC-decay constants $2\tau_{\mathrm{MC}}$ in single MRR simulations.
	The fitted constants are obtained by applying to the simulated responses the same MC calculation and decay-fitting procedure used for the experimental responses.
	}
	\begin{tabular}{lcccc}
		                              &                \multicolumn{2}{c}{Free carrier}                  & \multicolumn{2}{c}{Thermal}                        \\
		Parameter set                  & Physical $\tau$              & Fitted $2 \tau_{\mathrm{MC}}$ & Physical $\tau$                & Fitted $2 \tau_{\mathrm{MC}}$ \\
		\hline
		\textcite{vanvaerenbergh2012cascadable}  & \qty{5.3}{\nano\second} & \qty{6.8(2)}{\nano\second} & \qty{65}{\nano\second}  & \qty{74(4)}{\nano\second}  \\
		\textcite{borghi2021the}                 & \qty{40}{\nano\second}  & \qty{38(4)}{\nano\second}  & \qty{275}{\nano\second} & \qty{312(16)}{\nano\second} \\
	\end{tabular}
	\label{tab:mc-simulation-decays}
\end{table}

We next compare the MC-decay analysis with established relaxation-time estimation methods in terms of their required instrumentation and dependence on physical modeling.
However, we do not compare their reported numerical values because relaxation times depend on the fabrication process, die, device geometry, and operating conditions.
A pump--probe transient can directly probe free-carrier recombination.
\Textcite{borghi2021the} apply this method to a spiral waveguide on the same die as the MRR rather than to the resonator itself.
A short, intense pump pulse generates carriers through two-photon absorption, and the transient free-carrier absorption of a continuous-wave probe follows their recombination.
The inferred instantaneous carrier lifetime also depends on carrier concentration.
This approach therefore requires a dedicated pulsed-pump and probe path with time-resolved detection.

Thermal relaxation can be measured directly with a second optical probe.
\Textcite{gray2020thermo} scan a strong pump across one microring resonance to induce a thermo-optic bistability jump and monitor another longitudinal resonance with a weak fixed-wavelength probe.

Physical relaxation rates can also be inferred by fitting nonlinear dynamical models to measured waveforms or spectra.
\Textcite{johnson2006self} treat the free-carrier and thermal decay rates as free parameters in a coupled optical, carrier, and thermal model fitted to the self-pulsing waveform.
The remaining model quantities are constrained by low-power measurements, literature values, or finite-element calculations, so the inferred relaxation rates depend on the model and its independently specified parameters.
\Textcite{lin2008adiabatic} compare measured and simulated pulsed waveforms and spectra to infer an effective carrier lifetime, but they do not infer a thermal relaxation rate.

We therefore propose MC-decay analysis as a method for estimating effective relaxation times from the traces acquired for reservoir characterization.
The estimation requires neither additional instrumentation nor fitting a free-carrier--thermal model.
Under the exponential-response approximation tested in the single MRR simulations, the corresponding physical relaxation time is approximately twice the fitted MC-decay constant.

\subsection{Nonlinear Information-Processing Capacity}\label{sec:results-ipc}

The MC analysis establishes that the measured MRR network responses retain delayed input information, but it does not determine whether this information is available only through linear reconstruction or also through nonlinear transformations.
We therefore evaluate information-processing capacity (IPC) using the combined representation formed from all six measured MRR network output ports.
Below, we show that the result is predominantly linear, with a smaller quadratic single-delay contribution.

\version[A]{%
\Cref{fig:ipc-orders} shows that the linear capacity reaches approximately $C_1=\num{2.5(2)}$, while the quadratic capacity reaches approximately $C_2=\num{0.14(1)}$.
Orders 3--14 remain near zero, and \cref{fig:ipc-orders} displays only orders 1--5 because orders 6--14 are indistinguishable from zero at the plotted precision.
The box plots summarize the capacities across the investigated laser-frequency and mean-power operating points, and the highlighted marker for each order identifies the largest mean across the eight input series.
Since the first-order Legendre target is linear, $C_1$ is identical to MC under the same feature representation and normalization.
The much smaller $C_2$ therefore represents the additional capacity for reconstructing a quadratic function of one delayed input.
}
\version[B]{%
Across the investigated laser-frequency--power operating points, the largest mean linear and quadratic capacities are approximately $C_1=\num{2.5(2)}$ and $C_2=\num{0.14(1)}$, respectively.
The means and standard deviations are calculated across eight independent input series for each operating point.
Orders 3--14 remain near zero.
Since the first-order Legendre target is linear, $C_1$ is identical to MC under the same feature representation and normalization.
The much smaller $C_2$ therefore represents a limited additional capacity for reconstructing a quadratic function of one delayed input.
}

\version[A]{%
To determine how these contributions are distributed in time, \cref{fig:ipc-delays} resolves the capacity of each order by delay at the operating point that maximizes their sum.
The first-order profile $C_{1,k} \equiv \textrm{MC}_k$ nearly overlaps the summed capacity across the measured delay range, showing that linear reconstruction dominates at each delay.
Indeed, the quadratic profile $C_{2,k}$ has a similar decay shape but a much smaller amplitude, as shown more clearly in \cref{fig:ipc-quadratic}, while orders 3--14 remain near zero.
}
\version[B]{%
At the operating point that maximizes the summed delayed capacity, the first-order profile $C_{1,k}\equiv\mathrm{MC}_k$ nearly equals the sum over all evaluated orders throughout the measured delay range.
Linear reconstruction therefore dominates the delay-resolved capacity.
The quadratic profile $C_{2,k}$ has a similar decay shape but a much smaller amplitude.
}

These results provide an a posteriori check that linear MC captures the dominant delayed capacity among the evaluated single-delay targets.
They therefore support its use in the relaxation-scale analysis in \cref{sec:results-relaxation-times}.

\version[A]{%
\begin{figure}[htp]
	\includegraphics[width=\linewidth]{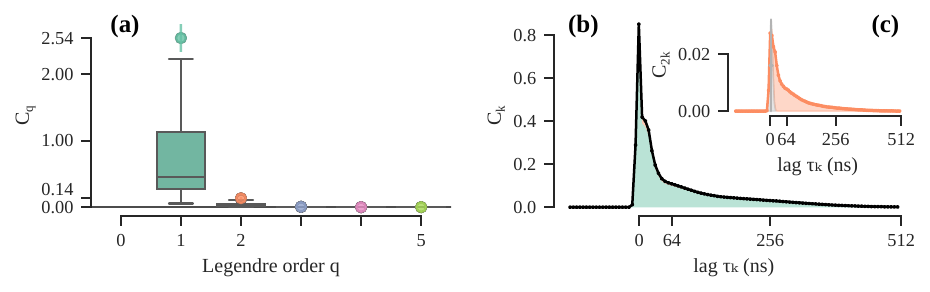}

	\begin{subcaptiongroup}
		\phantomcaption\label{fig:ipc-orders}
		\phantomcaption\label{fig:ipc-delays}
		\phantomcaption\label{fig:ipc-quadratic}
	\end{subcaptiongroup}

	\caption{%
		\textbf{Single-delay Legendre reconstruction capacities calculated at $T=\qty{10}{\nano\second}$ from the combined responses of MRR network ports 1, 2, 3, 5, 7, and 9.}
		Order $q=1$ is linear memory, but $q>1$ tests nonlinear functions of one delayed input.
		\subref{fig:ipc-orders} Delayed capacity $C_q$ for orders 1--5; orders 6--14 are omitted because they are indistinguishable from zero at the plotted precision.
		Colored box plots summarize the investigated laser-frequency--power operating points, and each colored marker and error bar gives the largest eight-series mean and its standard deviation; the displayed maxima are approximately $C_1=\num{2.5(2)}$ and $C_2=\num{0.14(1)}$.
		\subref{fig:ipc-delays} Delay-resolved capacities $C_{q,k}$ for each order $q$ at the operating point that maximizes the summed delayed capacity.
		The aqua $q=1$ contribution nearly overlaps the black curve, which is the sum over the evaluated orders $q=1,\ldots,14$.
		The smaller nonlinear contributions are not separately distinguishable at this scale.
		\subref{fig:ipc-quadratic} Enlarged view of the quadratic contribution $C_{2,k}$, shown in orange.
	}
	\label{fig:ipc}
\end{figure}
}
\subsection{NARMA Benchmark Task}\label{sec:results-narma}

The single-delay IPC analysis cannot determine whether the measured MRR network representation combines fading memory with nonlinear interactions between different input symbols.
We therefore evaluate NARMA reconstruction, which requires both properties within one task.
Following the definitions in \cref{sec:principle-narma}, we first evaluate the current-symbol condition and then increase the sample length and maximum symbol depth.
These two stages separate memory carried by the photonic response from output history supplied through digital storage.
NARMA5 with a symbol duration of \qty{40}{\nano\second} is the primary benchmark, and the subsequent comparisons vary the measured-response representation, temporal resolution, and NARMA order.

\subsubsection{NARMA Using the Current-Symbol Response}\label{sec:results-narma-tdrc}

We first test whether the current-symbol MRR network response contains enough input history to improve NARMA5 reconstruction without samples associated with preceding symbols.
At sample length $B=3$ and maximum symbol depth $D=0$, any information about preceding symbols must be encoded in the current photonic response.

The digital and out-of-resonance optical references in \cref{fig:narma-ports} give $\mathrm{NMSE}=\num{1.000}$ and $\mathrm{NMSE}=\num{1.000(3)}$, respectively, which are close to the error of a predictor that returns the target mean.
By comparison, the single-port resonant response gives $\mathrm{NMSE}=\num{0.696(2)}$, showing that the current photonic response already contains task-relevant information.
Adding selected MRR network outputs reduces the error further, and using all five selected output ports gives $\mathrm{NMSE}=\num{0.551(7)}$.
Because $D=0$, this improvement indicates that the current-symbol MRR network response contains information about preceding symbols.
The final value lies within the range obtained by the digital reference with maximum symbol depths from $D=1$ to 4, but this numerical correspondence does not imply storage of a specific number of input symbols.

The current-symbol representation still has higher NMSE than representations supplied with samples associated with preceding symbols.
This remaining gap motivates the sample-length expansion considered next.

The sample-window alignment and feature construction are detailed in \cref{sec:methods-narma-analysis}.

\subsubsection{Sample-Length and Response-Channel Expansion}\label{sec:results-narma-expansion}

We next determine whether the remaining NARMA improvement comes mainly from observing additional response channels or from increasing the maximum symbol depth.
We first compare the two forms of response-channel expansion.
The port-expanded representation in \cref{fig:narma-ports} reaches NMSE values of \num{0.551(7)}, \num{0.404(4)}, \num{0.130(3)}, and \num{0.117(4)} at sample lengths $B=3$, 9, 33, and 39, respectively.
The laser-frequency-expanded representation at port 1 in \cref{fig:narma-frequencies} gives \num{0.56(2)}, \num{0.42(2)}, \num{0.131(5)}, and \num{0.116(4)} at the same sample lengths.
Their close agreement at every sample length shows that output-port and laser-frequency expansion provide comparable performance for the selected representations.

We then compare the channel-expanded representations with a single resonant response and the two baselines.
At the same sample lengths, the single-port resonant representation gives NMSE values of \num{0.696(2)}, \num{0.544(5)}, \num{0.232(6)}, and \num{0.148(1)}, while the out-of-resonance optical baseline gives \num{1.000(3)}, \num{0.625(6)}, \num{0.28(8)}, and \num{0.159(7)}.
The digital baseline at the corresponding maximum symbol depths $D=0$, 1, 5, and 6 gives NMSE values of \num{1.000}, \num{0.642}, \num{0.196}, and \num{0.142}.
At $D=0$ and 1, the resonant single-port response has lower NMSE than both baselines, but this difference narrows as explicit symbol history is added.
In contrast, both channel-expanded representations remain below the single-port response and the two baselines at every evaluated maximum symbol depth.

Response-channel expansion therefore improves reconstruction, but increasing the maximum symbol depth produces the largest decrease in NMSE after the sample window begins to include samples associated with preceding symbols.
The readout then receives stored output samples from earlier symbols directly, so this decrease comes from explicit digital memory rather than longer intrinsic MRR network memory.
Sample lengths $B=33$ and 39 correspond approximately to maximum symbol depths $D=5$ and 6, respectively, and supply most of the temporal history required by the NARMA5 input term.
The mapping between sample length and maximum symbol depth is detailed in \cref{sec:methods-narma-analysis}.

Even at the same maximum symbol depth, the resonant representations have lower NMSE than the corresponding memoryless or digital references.
This residual difference could reflect nonlinear cross-symbol transformations in the MRR network response before the samples are stored.

Finally, we examine whether the best NARMA performance requires a resolved self-pulsing-like oscillation.
All laser-frequency--power operating points selected for the five-port representations lie outside the region where \cref{fig:fft-matrix-random-map} identifies self-pulsing-like responses.
A resolved self-pulsing-like component is therefore not required for the measured NARMA reconstruction.
Detailed operating-point comparisons, omitted here for simplicity, also show that responses with this component have higher NMSE under the investigated conditions.

\subsubsection{Dependence on Symbol Duration and Temporal Resolution}\label{sec:results-narma-resolution}

We next ask whether the lower NARMA error at $T=\qty{40}{\nano\second}$ results only from its greater number of measured samples per symbol.
To test this explanation, the temporal-resolution control averages non-overlapping groups of three consecutive samples from the \qty{40}{\nano\second} response while retaining approximately the same maximum symbol depth.
This averaging reduces the original sample lengths $B=3$, 9, 33, and 39 to 1, 3, 11, and 13 averaged features, respectively.

\Cref{fig:narma-timesteps} shows that the full-resolution \qty{40}{\nano\second} representation gives the lowest plotted mean NMSE.
Averaging causes only a small degradation, and the \qty{20}{\nano\second} representation performs worse than both \qty{40}{\nano\second} cases.
Temporal resolution therefore contributes to reconstruction but does not fully explain the difference between the two symbol durations.
Changing $T$ also changes the modulation rate relative to the carrier and thermal relaxation times, so the higher NMSE at \qty{20}{\nano\second} could reflect a different match between the input timing and the MRR network dynamics.

\subsubsection{Dependence on NARMA Order}\label{sec:results-narma-order}

\begin{figure}[htp]
	\includegraphics[width=\linewidth]{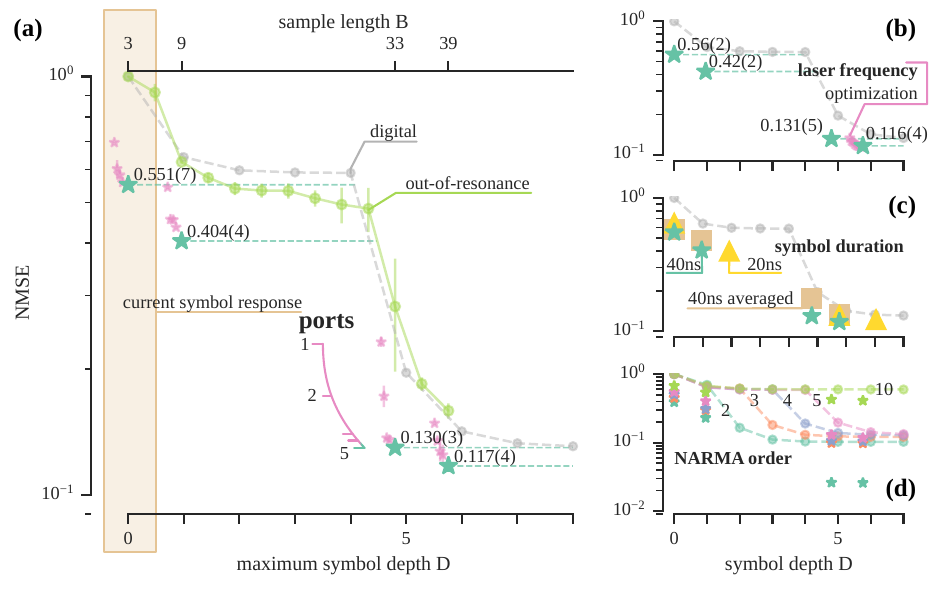}

	\begin{subcaptiongroup}
		\phantomcaption\label{fig:narma-ports}
		\phantomcaption\label{fig:narma-frequencies}
		\phantomcaption\label{fig:narma-timesteps}
		\phantomcaption\label{fig:narma-orders}
	\end{subcaptiongroup}

	\caption{%
		\textbf{NARMA reconstruction using digital inputs and measured MRR network responses with output-port, laser-frequency, and temporal feature expansion.}
		Markers and error bars give the mean and standard deviation over eight independent input series; gray circles joined by dashed lines denote references calculated directly from the programmed digital inputs, and colored markers denote linear-readout results calculated from measured optical responses.
		For $T=\qty{40}{\nano\second}$, sample lengths $B=3$, 9, 33, and 39 correspond to maximum symbol depths $D=0$, 1, 5, and 6, respectively.
		The tan region marks the current-symbol condition $D=0$.
		\subref{fig:narma-ports} NARMA5 reconstruction as output ports are added.
		Green circles denote the port-1 out-of-resonance reference, pink stars denote intermediate resonant port combinations, and aqua stars denote the final port set $\{1,3,5,7,9\}$.
		For $B=3$, 9, 33, and 39, the selected five-port representations use laser frequencies \qtylist{192.76;192.79;192.71;192.79}{\tera\hertz} and mean optical powers \qtylist{1.9;3.4;1.1;3.4}{\milli\watt}, respectively.
		\subref{fig:narma-frequencies} NARMA5 reconstruction as sequentially measured laser-frequency responses at port 1 are added; pink and aqua stars denote intermediate and selected combinations, respectively.
		\subref{fig:narma-timesteps} NARMA5 reconstruction from full-resolution \qty{40}{\nano\second} responses (aqua stars), three-sample-averaged \qty{40}{\nano\second} responses (tan squares), and full-resolution \qty{20}{\nano\second} responses (yellow triangles).
		\subref{fig:narma-orders} NARMA2, NARMA3, NARMA4, NARMA5, and NARMA10 evaluated with the same five-port set and sample lengths.
		Color-matched dashed circles and stars identify the digital references and measured optical results, respectively.
        The two aqua stars at low NMSE identify the NARMA2 experimental results for $B=33$ and 39.
	}
	\label{fig:narma}
\end{figure}

Finally, we test how reconstruction changes as the NARMA target requires longer temporal histories and more complex interactions among symbols.
\Cref{fig:narma-orders} compares the five-port photonic representation with the digital baseline for each NARMA order using the same sample lengths and corresponding maximum symbol depths.
The laser-frequency--power operating point is optimized independently for each NARMA order and sample length.
For example, the NARMA2 and NARMA10 results illustrate the change across the evaluated order range.
At maximum symbol depths $D=0$, 1, 5, and 6, the NARMA2 photonic representation gives NMSE values of \num{0.39(3)}, \num{0.23(2)}, \num{0.026(2)}, and \num{0.026(2)}, respectively, compared with \num{1.000}, \num{0.696}, \num{0.103}, and \num{0.103} for the digital baseline.
Thus, once the supplied history reaches $D=5$, the NARMA2 photonic error is already close to its minimum over the evaluated depths.
For NARMA10, the corresponding photonic NMSE values are \num{0.67(1)}, \num{0.544(8)}, \num{0.43(1)}, and \num{0.41(1)}, compared with \num{1.000}, \num{0.674}, \num{0.598}, and \num{0.597} for the digital baseline.
For the digital reference, we define the NARMA-$N$ knee as the rapid NMSE decrease to about \num{0.2} that occurs when the maximum symbol depth increases from approximately $D=N-1$ to $D=N$.
The evaluated NARMA10 depths end at $D=6$ and therefore do not reach its expected knee around $D=9$--10.
Even at $D=6$, the NARMA10 error remains above \num{0.4}.
The fixed measured representation therefore becomes less effective as the temporal and nonlinear requirements of the target increase.

The photonic representation has lower NMSE than the digital baseline for every evaluated NARMA order and maximum symbol depth.
However, the comparison is not feature-dimension matched.
The digital reference uses $D+1$ readout features, where $D$ is the maximum symbol depth, and the single-port out-of-resonance reference uses $B$ features, where $B$ is the number of consecutive output samples.
The resonant photonic representation uses $B$ times the number of combined output ports or laser frequencies.
For example, at $D=6$ and $B=39$, the five-port photonic representation uses 195 features, compared with 7 for the digital reference and 39 for the out-of-resonance reference.

Because the NARMA2 target contains a product of adjacent input symbols, its lower photonic NMSE is consistent with possible nonlinear cross-symbol processing.

Overall, the intrinsic MRR network memory profile is not well matched to the temporal dependencies of the evaluated NARMA tasks.
Supplying samples associated with preceding symbols through external digital memory addresses this limitation and produces most of the NMSE reduction.
Once this memory is available, the photonic representation provides an additional numerical improvement over the digital baseline, which could arise from nonlinear transformations or cross-symbol interactions in the MRR network response.
Because the NARMA improvement is dominated by the appropriate linear memory supplied externally, these results do not establish that the MRR network is suitable as a standalone reservoir for the evaluated NARMA tasks.

\section{Conclusion}\label{sec:conclusion}

This work characterizes how the free-carrier--thermal dynamics of a single MRR and a passive 64-MRR network retain and transform optical inputs modulated by random symbol sequences.
We compare constant and random excitation, map fading memory across laser-frequency--power operating points, combine responses measured at different laser frequencies and output ports, and test nonlinear processing through IPC and NARMA-$N$ reconstruction.
Together, these analyses connect observable temporal dynamics to the information that a linear readout can reconstruct from the measured photonic responses.

We found that, under random modulation, a dominant output-oscillation component associated with constant-drive self-pulsing remains observable mainly within the constant-drive self-pulsing region and persists across the investigated symbol durations.
However, the largest MC values occur near the boundary of this region rather than where the resolved oscillatory response is strongest.
Inside the self-pulsing region, the output retains a resolved slow oscillatory envelope, but both current-input reconstruction and delayed MC are low.
Operation in this region is therefore unfavorable for linear reconstruction of continuously modulated signals under the investigated conditions.
The memory landscapes of the single MRR and MRR network instead identify the self-pulsing boundary as the operating region where previous inputs are most accessible to a linear readout.

However, this conclusion is specific to fading-memory processing of continuously modulated inputs and does not imply that self-pulsing lacks computational utility.
In the non-fading regime demonstrated in our previous work, input perturbations encode information through either selection among self-pulsing attractors or phase shifts of the self-pulsing oscillation \cite{lugnan2025nonfading}.
This mode of computation stores information in a persistent self-pulsing state or phase rather than through a continuously decaying transient response.
Exploiting self-pulsing will therefore require a dedicated computational framework based on persistent-state dynamics rather than conventional fading-memory processing.
Such a framework should determine how reliably an MRR network can prepare, control, and read out self-pulsing states and phases for recurrent processing.

Another key finding in this work is that measurements at different laser frequencies and output ports access different resonance detunings and optical paths, providing complementary observations of the MRR network dynamics.
Combining these responses increases the delayed input information accessible to the readout and provides a broader measured representation.
This post-acquisition combination changes the observable representation but not the physical dynamics that generate each response.

Also, the MC profiles contain fast and slow decay components associated mainly with free-carrier and thermal relaxation.
Fitting these components provides a less experimentally demanding method for estimating effective timescales of nonlinear MRR dynamics from the same random-input measurements.

For Legendre functions of one delayed input, IPC shows that the measured MRR network representation contains predominantly linear delayed input information and a smaller quadratic contribution.

NARMA-$N$ reconstruction shows that the current-symbol MRR network response contains task-relevant information about preceding symbols, but its intrinsic memory profile is not sufficient for the evaluated tasks.
Increasing the maximum symbol depth through digital storage supplies the missing history and forms a hybrid photonic--digital processor.
Under the investigated conditions, the passive MRR network by itself is therefore not an effective standalone reservoir for the evaluated NARMA tasks.

Previous MRR-based RC results concern different memory resources and readout constructions.
For example, numerical NARMA10 implementations combine temporal virtual nodes with external optical feedback or high-Q auxiliary-ring storage \cite{donati2021microring,gironcastro2024effects,ren2024photonic}.
Intrinsic-memory demonstrations address short delayed binary tasks or compact classification \cite{borghi2021reservoir,bazzanella2022microring}, and some successful task conditions rely primarily on photodetection for nonlinearity \cite{donati2021microring,donati2024time}.
Here, we characterize linear memory, single-delay nonlinear capacity, and NARMA performance using port- and laser-frequency-diverse observations, without external optical feedback.
The observed limitation therefore concerns the measured passive configuration and does not exclude successful MRR-based processing with differently engineered memory, input timing, or readout representations.

These findings suggest that task-specific fading memory profiles require control over the relaxation dynamics of individual resonators rather than only an increase in the number of passive MRRs.
For example, integrated p--n junctions could provide active carrier extraction and thereby tune the fast free-carrier relaxation scale of each resonator.
Assigning different carrier-relaxation times to individual resonators could provide responses that emphasize different portions of the input history.
Faster responses would emphasize recent inputs, while slower responses could retain information from earlier inputs.
Combining these responses could therefore shape the accessible fading-memory profile to better match the temporal requirements of a target task.
However, shaping the slower thermal contribution would require an additional thermal or structural control strategy.
Future experiments should test whether such active relaxation-scale engineering changes the MC profile and improves tasks whose temporal requirements match the designed profile.

\section{Methods}\label{sec:methods}

\begin{figure}[htp]
	\includegraphics[width=\linewidth]{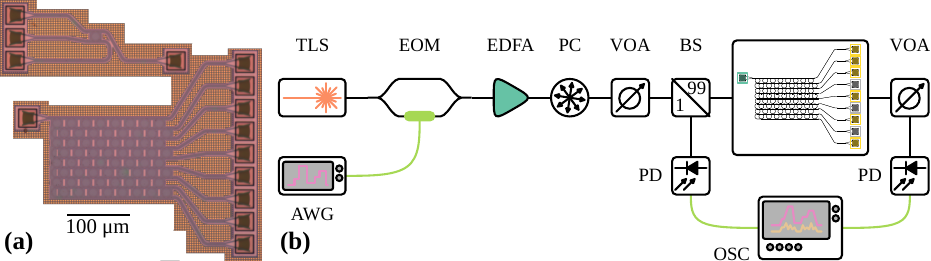}

	\begin{subcaptiongroup}
		\phantomcaption\label{fig:methods-picture}
		\phantomcaption\label{fig:methods-setup}
	\end{subcaptiongroup}

	\caption{%
		\textbf{Photonic integrated circuit and experimental setup.}
		\subref{fig:methods-picture}  Photograph of the photonic integrated circuit taken through an optical microscope, showing the single MRR in an add-drop configuration at the top and the $8\times8$ MRR network at the bottom; scale bar, \qty{100}{\micro\meter}.
		\subref{fig:methods-setup} Optical and electrical measurement setup.
		The arbitrary waveform generator (AWG) drives the electro-optic modulator (EOM), which intensity-modulates the tunable laser source (TLS).
		The optical signal then passes through the erbium-doped fiber amplifier (EDFA), polarization controller (PC), variable optical attenuator (VOA), and 99:1 beam splitter (BS).
		The BS directs approximately \qty{1}{\percent} of the optical power to the input-monitor photodetector (PD) and the remainder to the photonic circuit; a second VOA precedes the output PD, and the oscilloscope (OSC) records both detector signals.
		Black and green lines identify optical and electrical connections, respectively; yellow and gray frames identify the six measured and three excluded MRR network outputs.
	}
	\label{fig:methods}
\end{figure}

\subsection{Photonic Integrated Circuit}\label{sec:methods-circuit}

The photonic integrated circuit (PIC) is fabricated by imec (Leuven, Belgium) on a silicon-on-insulator (SOI) platform.
The silicon waveguides have a cross-section of $\qty{450}{\nano\meter}\times\qty{220}{\nano\meter}$ and are embedded in silica cladding.
The PIC shown in \cref{fig:methods-picture} contains one reference microring resonator, termed the single MRR, and one MRR network.
The two devices are positioned close together on the same chip \cite{lugnan2025nonfading,foradori2026timeseries}.
Each input and output grating coupler has a nominal insertion loss of approximately \qty{3}{\decibel}.
Unless otherwise stated, the reported mean optical powers are on-chip input powers referenced immediately after the input grating coupler.

\subsubsection{Single MRR}\label{sec:methods-isolated-mrr}

The single add-drop MRR has a racetrack geometry with circular sections of bend radius \qty{7}{\micro\meter} and straight coupling sections of length \qty{0.71}{\micro\meter}.
The gap between the resonator and each bus waveguide is \qty{0.2}{\micro\meter}.
The two bus waveguides provide access to the input, through, and drop ports shown in \cref{fig:intro-ring}.
All measurements of the single MRR reported in this work are acquired at the through port.

\subsubsection{MRR Network}\label{sec:methods-mrr-matrix}

The MRR network consists of an $8\times8$ arrangement of 64 passive MRRs connected by bus waveguides, as shown in \cref{fig:intro-matrix}.
Every MRR in the network uses the racetrack geometry described for the single MRR in \cref{sec:methods-isolated-mrr}.
Adjacent MRRs in the same row are separated horizontally by \qty{22.7}{\micro\meter}, and the resonators in neighboring rows are horizontally offset by \qty{11.35}{\micro\meter}.

The MRR network has one physical input on the left and nine physical outputs on the right.
Output ports are numbered from top to bottom.
Ports 1, 2, 3, 5, 7, and 9 are used in the experiments reported here.
Ports 4, 6, and 8 are excluded from the time-resolved analyses because the output fiber could not be aligned reliably to their grating couplers.
The MRR network is fully passive, and no individual MRR is actively tuned after fabrication.
Active tuning therefore does not compensate for fabrication-induced differences among the MRR resonance frequencies.

\subsection{Experimental Setup}\label{sec:methods-setup}

\subsubsection{Optical Excitation and Coupling}\label{sec:methods-optical-excitation}

\Cref{fig:methods-setup} shows the optical and electrical connections used for excitation and acquisition.
A continuous-wave tunable laser source (Pure Photonics) generates the optical carrier in the telecommunications C-band around \qty{1550}{\nano\meter}.
The laser operates at a selected laser frequency and constant output power.
An arbitrary waveform generator (AWG; Spectrum DN2.663-02) drives a lithium-niobate electro-optic modulator (iXblue MXAN-LN-10) that intensity-modulates the light.
\Cref{sec:methods-input} details the generation of the electrical waveforms.

An erbium-doped fiber amplifier (Thorlabs) amplifies the optical signal after modulation.
A polarization controller aligns the input polarization with the polarization supported by the grating couplers.
An electronically controlled variable optical attenuator (VOA) sets the launched optical power.
A 99:1 fiber splitter directs approximately \qty{1}{\percent} of the signal to the input-power monitor and the remaining light to the chip.

The input and output fibers are mounted on three-axis piezoelectric translation stages.
Their cleaved facets are inclined by approximately \qty{10}{\degree} relative to the chip normal and aligned with the selected grating couplers.
The output fiber collects light transmitted by the selected circuit output and directs it through a second variable optical attenuator before photodetection.
This attenuator protects the fast detector from excessive optical power.

A proportional--integral--derivative controller and a \qty{10}{\kilo\ohm} thermistor regulate a Peltier element that stabilizes the chip temperature.

\subsubsection{Photodetection and Electronic Instrumentation}\label{sec:methods-detection}

A slow photodetector with a \qty{30}{\kilo\hertz} bandwidth (New Focus M2033) monitors the \qty{1}{\percent} output of the fiber splitter.
Its electrical signal provides the reference used to set and monitor the mean optical input power delivered to the chip.
A \qty{600}{\mega\hertz} photodetector (Menlo Systems FPD610-FC-NIR) converts the transmitted time-dependent response into an electrical signal.

A PicoScope 6000-series oscilloscope records the electrical signals from the input-power monitor and fast photodetector while the AWG repeatedly generates the electrical modulation waveform.
Output sampling and input--output synchronization are described in \cref{sec:methods-timing}.
The output attenuator is adjusted to prevent fast-photodetector saturation and keep the recorded signal within the oscilloscope input range.

\subsection{Input-Waveform Generation}\label{sec:methods-input}

\subsubsection{Random and Constant Excitation}\label{sec:methods-input-random-constant}
Each random input comprises \num{100000} independent values drawn uniformly from 256 discrete levels, $u_n\in\{0,\ldots,255\}$.
Eight independent random realizations are used.
The same digital sequence is replayed for a given realization when the laser frequency, mean input power, output port, or device changes.
Replaying the same input realizations enables alignment of separately acquired responses with the corresponding input symbols and comparison at matched sequence positions during post-processing.

The digital levels are mapped to target optical powers between zero and $P_{\max}$.
For an equiprobable 256-level sequence, the expected target mean optical power is $P_{\max}/2$.
Constant-excitation measurements use an unmodulated optical input set nominally to the measured mean power of the corresponding random waveform.
However, because the AWG output differs between modulated and constant operation, the measured mean power of each random waveform is approximately \qty{90}{\percent} of the corresponding constant-input power at the same nominal setting.
The random- and constant-excitation responses are therefore compared at nominally matched, but not exactly equal, measured mean optical powers.

\subsubsection{AWG Encoding and Symbol Duration}\label{sec:methods-awg-encoding}

The AWG converts the target sequence into the electrical waveform driving the electro-optic modulator.
The selected AWG operating mode uses waveform points separated by \qty{10}{\nano\second}.
The AWG generates longer symbol durations by repeating each target value on this grid.
Calibrated voltage levels map the target sequence to optical powers between zero and $P_{\max}$.

For $T=\qty{10}{\nano\second}$, each random value is represented by one AWG point.
The AWG linearly interpolates between consecutive points.
The optical modulation therefore changes continuously between adjacent random levels and has piecewise-linear transitions without a constant-level plateau.
For $T=\qty{20}{\nano\second}$ and $T=\qty{40}{\nano\second}$, each target value is repeated two and four times, respectively.
Each transition between successive levels remains confined to one \qty{10}{\nano\second} interval.
The repeated points produce plateaus whose duration increases with $T$.

The AWG waveform is organized as a continuously repeated acquisition cycle.
Each cycle begins with a marker pulse, followed by a pause, a constant-input segment, and a second pause.
The cycle then contains one block for each tested symbol duration, $T=\qtylist{10;20;40}{\nano\second}$.
Each block contains eight independent random sequences grouped into four pairs, with a pause after each pair.
A second constant-input segment follows the final pause after all three symbol-duration blocks.
The three blocks contain 24 random sequences in total.
A final marker pulse and a padding interval complete the cycle before repetition.
The marker pulses identify the cycle boundaries during post-processing.
The pauses separate consecutive signal groups and are intended to reduce memory carryover.
Each constant and random segment begins and ends with a smooth power ramp to avoid abrupt changes in the optical drive.
The subsequent analysis excludes the ramp, marker, pause, and padding intervals.

\subsection{Data Acquisition}\label{sec:methods-acquisition}

\subsubsection{Acquisition Procedure and Operating-Point Sweeps}\label{sec:methods-operating-sweeps}

The AWG continuously repeats the acquisition cycle described in \cref{sec:methods-awg-encoding}.
The oscilloscope records the fast-photodetector output for each fixed combination of device, output port, laser frequency, and VOA setting during application of this waveform.
Tuning the laser frequency and allowing it to settle take longer than adjusting the VOA.
The laser frequency therefore defines the outer acquisition loop, and the mean input power defines the inner loop.
The experiment steps the VOA through the device-specific power grid at each laser frequency before advancing the laser to the next selected laser frequency.
The same AWG cycle contains the constant and random segments, so both are acquired under the same laser-frequency, VOA, and optical-path conditions.
The nonlinear sweep for the single MRR contains 21 equally spaced laser frequencies from \qty{192.60}{\tera\hertz} to \qty{192.70}{\tera\hertz}, corresponding to a \qty{5}{\giga\hertz} increment.
These measurements are acquired at the through port.
The single-MRR constant-excitation measurements include a low-power point at \qty{0.01}{\milli\watt} and a power sweep from \qty{0.3}{\milli\watt} to \qty{7.3}{\milli\watt} in increments of approximately \qty{0.58}{\milli\watt}.
The MRR network is measured at 14 equally spaced laser frequencies from \qty{192.68}{\tera\hertz} to \qty{192.86}{\tera\hertz}, including both endpoints and corresponding to an approximately \qty{13.8}{\giga\hertz} increment.
The MRR-network constant-excitation power sweep extends from \qty{1.2}{\milli\watt} to \qty{12.3}{\milli\watt} in increments of approximately \qty{0.85}{\milli\watt}.
For both devices, the random-excitation mean optical power at each nominal setting is approximately \qty{90}{\percent} of the corresponding constant-excitation value because of the modulation-dependent AWG output described in \cref{sec:methods-input-random-constant}.
Each laser frequency is measured in a separate acquisition.
Post-processing assembles representations that combine responses measured at different laser frequencies from the resulting traces.

\subsubsection{Out-of-Resonance Acquisition}\label{sec:methods-reference-acquisition}

The out-of-resonance references are special-case acquisitions outside the laser-frequency--power sweeps used to characterize the nonlinear resonant response.
The out-of-resonance reference for the single MRR is measured at the through port at a laser frequency of \qty{192.55}{\tera\hertz}.
The out-of-resonance reference for the MRR network is measured at port 1 at a laser frequency of \qty{192.60}{\tera\hertz}.
For each device, the reference is acquired at the maximum input power of the corresponding sweep to increase the detected signal-to-noise ratio.
The same repeated AWG cycle is applied under these reference conditions.

\subsubsection{Port and Device Reconfiguration}\label{sec:methods-diverse-acquisition}

The experiment uses one fast detection channel and therefore acquires the six MRR network outputs sequentially.
The experiment completes the laser-frequency--power sweep at one output before aligning the output fiber to the next measured port in the sequence 1, 2, 3, 5, 7, and 9.
The experiment then repeats the complete operating-point sweep with the same AWG waveform and random input realizations.
The experiment therefore does not demonstrate simultaneous multiport detection.
Simultaneous multiport acquisition would require parallel detection and acquisition channels for all measured outputs.

Device reconfiguration between the single MRR and the MRR network requires alignment of both the input and output fibers to the corresponding grating couplers.
The experiment repeats the complete acquisition procedure with the device-specific laser-frequency and power grids after this reconfiguration.
The sequential port and device acquisitions use no intermediate recalibration of laser frequency or optical power.

\subsubsection{Timing, Sampling, and Trace Alignment}\label{sec:methods-timing}

The oscilloscope samples the fast-photodetector output at intervals of $\Delta t=\qty{6.4}{\nano\second}$.
The AWG and oscilloscope operate on different time grids and are not synchronized sample by sample.
The marker pulses embedded in each repeated AWG cycle define the cycle boundaries during post-processing.
Post-processing excludes the marker, pause, ramp, and padding intervals before extracting the random and constant-input responses.
Post-processing then aligns the extracted input and output traces in time.
\Cref{sec:methods-mc-analysis} describes the resampling procedures used for MC and IPC.
The analysis excludes acquisitions that cannot be parsed or aligned reliably, and the operating-point maps display these acquisitions as empty cells.

\subsection{Data Analysis}\label{sec:methods-analysis}

The following subsections describe implementation details that supplement the definitions and analysis framework introduced in \cref{sec:characterization-metrics}.
The experimental data analysis uses Python.
The analysis uses PyTorch for CUDA-accelerated tensor operations and the \texttt{torch.linalg} module for linear-algebra calculations \cite{paszke2019pytorch}.

\subsubsection{Spectral Analysis and Classification}\label{sec:methods-spectral-analysis}

Spectral analysis is applied independently to each acquired output trace.
Let $x_j$ denote the uniformly sampled temporal response after the selected preprocessing.
A fast Fourier transform of $x_j$ gives its discrete Fourier transform $X_m$.
The unnormalized periodogram is defined as follows.
\begin{equation}
	\mathrm{PSD}_m=\abs{X_m}^2.
\end{equation}
For random excitation, the periodogram is logarithmically transformed.
\begin{equation}
	L_m=\log\!\left(\mathrm{PSD}_m\right).
\end{equation}
The log PSD is smoothed to reduce bin-to-bin fluctuations.
The temporal Fourier-frequency bin at the maximum of the smoothed log PSD defines the dominant output-oscillation frequency.
The acquisition duration sets the temporal Fourier-frequency resolution.

The normalization uses log-PSD values from \qty{10}{\mega\hertz} to \qty{12.5}{\mega\hertz} to calculate a reference mean and standard deviation.
It subtracts the reference mean from the full log PSD and divides the mean-centered values by the reference standard deviation.

The signed Pearson correlation $r$ between smoothed spectra quantifies the similarity of responses measured at the same operating point with $T=\qty{10}{\nano\second}$ and $T=\qtylist{20;40}{\nano\second}$.

The analysis assigns empirical response labels for random excitation using the area $A$ under the normalized output log PSD and the error $E$ of an exponential-envelope fit.
Approximately linear, input-like responses satisfy $A\leq T_A$.
Responses with $A>T_A$ and $E\leq T_E$ are labeled nonlinear without a resolved self-pulsing signature.
Responses with $A>T_A$ and $E>T_E$ are labeled self-pulsing-like.
An operating point is assigned the self-pulsing-like map label only when the resolved signature is present in all eight random-input traces.
For responses assigned the self-pulsing-like label, the reported dominant output-oscillation frequency is the temporal-frequency bin at the maximum of the smoothed log PSD.
These empirical labels summarize the measured spectral features but do not independently establish distinct physical dynamical regimes.
The thresholds $T_A$ and $T_E$ are selected automatically from the corresponding single MRR metric distributions.

For constant excitation, each trace is mean-subtracted and variance-normalized before the FFT, with no additional preprocessing.
The classification metric is then $S=\sum_f\abs{\mathrm{FFT}(x)(f)}$.
Responses satisfying $S<T_{\mathrm{SP}}$ are labeled self-pulsing.
Responses satisfying $S\geq T_{\mathrm{SP}}$ are labeled nonperiodic.
The nonperiodic label denotes the absence of a resolved periodic self-pulsing signature under this metric and may include either linear or nonlinear responses.
$T_{\mathrm{SP}}$ is selected automatically from the corresponding single MRR metric distribution.

The single-MRR random--constant comparison accounts for their different measured mean powers by locating, at the same laser frequency, the two adjacent constant-power states that enclose each random-modulation power.
The upper constant-power state must be self-pulsing for the state to enter the calibration-matched comparison.
When the lower state is also self-pulsing, the constant-drive self-pulsing frequency is linearly interpolated in power between the two bounds.
The signed frequency difference is defined as $\Delta f=f_{\mathrm{random}}-f_{\mathrm{constant,interp}}$, and the relative difference is $\Delta f/f_{\mathrm{random}}$.
The aggregate comparisons report the mean absolute differences $\abs{\Delta f}$ and $\abs{\Delta f/f_{\mathrm{random}}}$ so that positive and negative deviations do not cancel.

The MRR network metric distributions are strongly class-imbalanced and do not support stable independent threshold estimation.
The single MRR thresholds are therefore transferred unchanged to the MRR network.
An a posteriori check indicates that the transferred thresholds are also consistent with the corresponding class separation in the MRR network data.

\subsubsection{Memory-Capacity Analysis}\label{sec:methods-mc-analysis}

The reported MC analysis uses responses acquired with $T=\qty{10}{\nano\second}$.
The input and output use different time grids and are not synchronized sample by sample.
The analysis linearly interpolates the input, samples it at the output timestamps, and aligns it with the measured response.
The resulting common grid has an interval $\Delta t=\qty{6.4}{\nano\second}$.
The analysis evaluates negative delays as a causality and alignment check.
Their reconstruction targets are future input values relative to the measured response.
The corresponding capacities are expected to be zero apart from finite-data fluctuations.
Positive delays extend to $\tau_k=\qty{512}{\nano\second}$.

For each lag $k$, an affine linear readout reconstructs the delayed input from the photonic feature vector,
\begin{equation}
	\widehat{u}_{n-k}=b_k+\mathbf{w}_k^{\mathsf{T}}\mathbf{x}_n.
\end{equation}
The reconstruction residual is
\begin{equation}
	e_{n,k}=u_{n-k}-\widehat{u}_{n-k}.
\end{equation}
The raw delay-resolved memory coefficient is the corresponding coefficient of determination,
\begin{equation}
	\mathrm{MC}_{k,\mathrm{raw}}
	=
	1-\frac{\mathrm{SSE}_k}{\mathrm{SST}_k},
\end{equation}
where $\mathrm{SSE}_k$ is the sum of squared errors and $\mathrm{SST}_k$ is the total sum of squares:
\begin{equation}
	\mathrm{SSE}_k=\sum_n e_{n,k}^2,
	\qquad
	\mathrm{SST}_k=\sum_n\left(u_{n-k}-\overline{u}_k\right)^2,
\end{equation}
where $\overline{u}_k$ is the mean of the delayed target over the samples included in the evaluation.
For MC and IPC, each readout is fitted and evaluated on the same aligned samples, so these capacities describe in-sample reconstruction.
The affine model includes the constant mean predictor as a special case, so the least-squares solution satisfies $\mathrm{SSE}_k\leq\mathrm{SST}_k$.
Consequently, $\mathrm{MC}_{k,\mathrm{raw}}$ is nonnegative by construction and equals zero when the optimal affine fit does not improve on the mean predictor.

The finite-width instantaneous peak is fitted on the dimensionless lag grid with a Gaussian.
\begin{equation}
	G(k)=A\exp\!\left[-\frac{(k-k_0)^2}{2\sigma_k^2}\right].
\end{equation}
$A$ is the peak amplitude.
$k_0$ provides a fine adjustment of the temporal alignment in sample units.
$\sigma_k$ is the Gaussian width in lag samples.
The corresponding physical offset and width are $\tau_0=k_0\Delta t$ and $\sigma_\tau=\sigma_k\Delta t$, respectively.

A discrete fast Fourier transform is applied to each time series.
The analysis multiplies the Fourier-domain representation by the phase factor corresponding to the fitted shift $k_0$.
An inverse transform returns the shifted series to the time domain and places the fitted instantaneous peak at $k=0$.

For every single-response representation, the MC calculated from the shifted series agrees with the MC calculated from the original series.
This agreement indicates that the alignment correction does not alter the measured single-response capacity.
The correction improves multirepresentation alignment.

For each input symbol duration $T$, the fitted $\sigma_k$ values are compatible across output responses.
The corresponding common fitted width increases linearly with $T$.
The instantaneous contribution $R_{\mathrm{raw}}^2$ is the Gaussian area on the dimensionless lag grid.
\begin{equation}
	R_{\mathrm{raw}}^2=\int_{-\infty}^{\infty}G(\kappa)\,\mathrm{d}\kappa
	=A\sigma_k\sqrt{2\pi},
\end{equation}
where $\kappa$ is the continuous coordinate corresponding to the discrete lag index $k$.
The area is therefore defined on a dimensionless lag coordinate and is not integrated over physical time.
The same fitting and area calculation applied to the system-specific out-of-resonance response defines $R_{\mathrm{off}}^2$.
The complete raw delay-resolved profile is normalized as
\begin{equation}
	\mathrm{MC}_k
	=
	\frac{\mathrm{MC}_{k,\mathrm{raw}}}{R_{\mathrm{off}}^2}.
\end{equation}
This normalization assigns unit integrated weight to the broadened instantaneous response of the out-of-resonance reference.
It compensates for the redistribution of each input-delay contribution over neighboring lag points under the assumption that the acquisition-induced broadening is approximately independent of delay.
It does not remove noise, residual optical dynamics, or deviations from the assumption that the out-of-resonance response is linear and memoryless.

For a resonant response, the normalized instantaneous contribution is
\begin{equation}
	R^2
	=
	\frac{R_{\mathrm{raw}}^2}{R_{\mathrm{off}}^2}.
\end{equation}
Negative-lag terms vanish by causality apart from finite-data fluctuations and provide a causality and alignment check.
The finite-range delayed memory capacity is obtained by subtracting the complete instantaneous contribution from the normalized delay-resolved profile:
\begin{equation}
\mathrm{MC}
=\sum_k\mathrm{MC}_k-R^2.
\end{equation}

The operating-point maps use a single Gaussian Mixture Model (GMM) fitted jointly to the single MRR and port-1 MRR network measurements.
The fit uses normalized $R^2$ and total MC.
The resulting labels are displayed separately on the two maps.
They identify empirical groups with similar reconstruction properties.

\subsubsection{Memory-Capacity Feature Combination}\label{sec:methods-mc-feature-combination}

The analysis evaluates memory capacity for eight independent random input series and refits the linear readout separately for each series and photonic configuration.
This procedure gives eight MC values per configuration.
Their mean ranks the configurations, and their standard deviation quantifies variation across the input series.

\paragraph{Greedy Frequency Combination}\label{sec:methods-mc-greedy-frequency}
Each candidate is an experimentally measured response at one laser frequency and one mean input power.
The algorithm first selects the laser-frequency--power pair with the largest mean MC across the eight series.
The algorithm removes all other powers measured at each selected laser frequency, so it retains at most one power per laser frequency.
At each subsequent step, the algorithm provisionally concatenates each remaining candidate response with the responses already selected.
The algorithm refits the readout for all eight series for each provisional combination.
It retains the candidate with the largest resulting mean MC.
The single MRR analysis selects 14 of the 21 available laser frequencies.
The port-1 MRR network analysis selects all 14 available laser frequencies.

\paragraph{Port Combination}\label{sec:methods-mc-simple-ports}

For each laser frequency, mean input power, and independent input series, all nonempty combinations of the six sequentially measured outputs of the MRR network are considered.
The response traces from the ports in each combination are concatenated into a joint feature representation.
A separate linear readout is fitted for each port combination.

\subsubsection{Memory-Decay Fitting}\label{sec:methods-mc-decay-fits}

The analysis fits the fast and slow memory components independently for each of the eight random input series.
The corresponding fitted time constants are interpreted as effective input--output decay constants.
Each component is modeled by an exponential function.
\begin{equation}
	\mathrm{MC}_k=A\exp(-\tau_k/\tau_{\mathrm{MC}}).
\end{equation}
A logarithmic transformation is applied to $\mathrm{MC}_k$.
\begin{equation}
	\log(\mathrm{MC}_k)=\log(A)-\tau_k/\tau_{\mathrm{MC}}.
\end{equation}
A weighted linear regression fits $\log(\mathrm{MC}_k)$ as a function of physical delay, and the fitted slope determines $\tau_{\mathrm{MC}}$.
The analysis fits the fast and slow components over separate delay intervals.

The procedure selects the fast and slow fit intervals independently for each input series.
Predefined minimum and maximum lengths constrain the number of consecutive points in each candidate interval.
The fast-fit interval starts at $k=0$.
The fitted Gaussian instantaneous contribution is not subtracted from the delay-resolved profile before the exponential fit, so the fast interval includes the broadened instantaneous peak.
The procedure evaluates each permitted fast-interval length and retains the interval that minimizes the weighted regression-error criterion.
Candidate slow-fit intervals start after the final point of the selected fast interval.
The procedure varies both the starting point and length of the slow interval and retains the candidate that minimizes the same error criterion.
The two selected non-overlapping intervals define the fast and slow exponential fits.

\subsubsection{Single MRR Simulation Validation}\label{sec:methods-mc-simulation}

Numerical simulations of the single MRR test the memory-decay analysis under controlled model conditions.
The test uses the dynamical models reported by \textcite{vanvaerenbergh2012cascadable} and \textcite{borghi2021the} with their respective parameter sets.
In the model of \textcite{borghi2021the}, the free-carrier lifetime depends on the free-carrier concentration.
Here, we make one modification by fixing this lifetime at its reported mean value of \qty{40}{\nano\second}.
Each simulation is driven by a random sequence containing 32 discrete input levels and on the order of $10^3$ symbols.
The output is sampled at four points per input symbol, corresponding to an interval of $T/4$.
The simulated input and output share a common time grid, so input alignment is unnecessary.
Responses simulated at multiple laser frequencies are combined using the same greedy laser-frequency-combination algorithm applied to the experimental data.
The resulting memory profiles are processed using the same automatic fast and slow decay-fitting procedure as the experimental profiles.
The simulation software and its analysis pipeline are implemented in Julia \cite{bezanson2017julia}.

\subsubsection{Information-Processing-Capacity Analysis}\label{sec:methods-ipc-analysis}

The IPC analysis uses the same $T=\qty{10}{\nano\second}$ responses as the MC analysis.
The interpolated and resampled input is mapped linearly to $z_n\in[-1,1]$.
For polynomial order $q\geq1$ and delay $k$, the reconstruction target is $P_q(z_{n-k})$, where $P_q$ is the Legendre polynomial of order $q$.
Order 0 is excluded because its target is constant, which gives zero variance and an undefined coefficient of determination.
The evaluated orders are $q=1, 2, \ldots, 14$.

For each order and delay, the analysis uses the same affine linear readout and coefficient-of-determination calculation as the MC analysis in \cref{sec:methods-mc-analysis}.
Dividing this coefficient by $R_{\mathrm{off}}^2$ gives the normalized delay-resolved capacity $C_{q,k}$ and applies the same correction for acquisition-induced broadening as in MC.
Interpolation and resampling modify the input distribution, so the evaluated Legendre targets are not necessarily exactly orthogonal.
The capacities are therefore interpreted as descriptive order-resolved contributions.

A Gaussian is fitted to the raw instantaneous profile for each order.
Its fitted area divided by $R_{\mathrm{off}}^2$ defines the normalized instantaneous contribution $R_q^2$.
The delayed capacity for order $q$ over the finite measured lag range is
\begin{equation}
	C_q
	=\sum_k C_{q,k}-R_q^2.
\end{equation}
The linear target gives $C_{1,k}=\mathrm{MC}_k$ and $C_1=\mathrm{MC}$.

The readout uses the concatenated responses from all six measured outputs of the MRR network.
Capacities are evaluated independently for the eight input series, and each operating-point mean is calculated across these series.

\subsubsection{NARMA Analysis}\label{sec:methods-narma-analysis}

\Textcite{atiya2000new} introduced the tenth-order nonlinear recurrence now known as NARMA10.
We write its order-$N$ generalization in parametric form as
\begin{equation}
	y_{n+1}
	=\alpha y_n
	+\beta y_n\sum_{i=0}^{N-1}y_{n-i}
	+\gamma u_{n-N+1}u_n
	+\delta.
\end{equation}
The coefficient $\alpha$ weights linear feedback from the current target value, and $\beta$ weights its nonlinear interaction with the previous $N$ target values.
The input term multiplies the current input $u_n$ by the input $N-1$ symbols earlier, $u_{n-N+1}$, with coefficient $\gamma$.
The coefficient $\delta$ provides a constant offset.
For $N=10$, the input term is therefore $u_{n-9}u_n$; together with $\alpha=0.3$, $\beta=0.05$, $\gamma=1.5$, $\delta=0.1$, and $u_n\in[0,0.5]$, this gives the NARMA10 recurrence used by \textcite{atiya2000new}.

The same work introduced a separate second-order nonlinear system.
\Textcite{rodan2011minimum} later referred to that system as NARMA2; see also \cite{wringe2025reservoir}.
In this work, NARMA2 instead denotes the $N=2$ instance of the generalized recurrence above, for which the input term is $u_{n-1}u_n$.

The target used here applies a hyperbolic tangent to the complete order-$N$ update:
\begin{equation}
	y_{n+1}=\tanh\!\left(
	\alpha y_n
	+\beta y_n\sum_{i=0}^{N-1}y_{n-i}
	+\gamma u_{n-N+1}u_n
	+\delta
	\right).
\end{equation}
We use the same coefficients and input range for every evaluated order $N$.
The $\tanh$ function is approximately linear over most of the range used here and bounds every target to $(-1,1)$.
This modification avoids divergent target trajectories under the common coefficient and input convention used here \cite{wringe2025reservoir}.
The recurrence is initialized with $y_n=0$ and $u_n=0$ for indices preceding the available sequence.
The evaluated orders are $N=2$, 3, 4, 5, and 10, with NARMA5 as the primary benchmark.

The first acquired output sample whose timestamp falls within the first input-symbol interval defines the alignment anchor.
Output timestamps are assigned to input-symbol intervals relative to this anchor.
$T/\Delta t$ is not an integer, so the input-symbol boundaries do not coincide with the output-sampling instants.
The relative alignment between the symbols and output samples therefore shifts along the sequence.
Unlike the MC analysis, the NARMA analysis does not interpolate and resample the input at the output timestamps.
It retains the original symbol-indexed input sequence and one NARMA target value for each symbol.
Because the AWG linearly interpolates between programmed symbol levels, resampling would mix neighboring symbols and introduce interpolation-induced information leakage into the input representation.
Associating multiple output samples with each original symbol also preserves the output dynamics induced within the symbol interval for the reconstruction test.

Symbol durations of \qty{20}{\nano\second} and \qty{40}{\nano\second} are considered.
With the \qty{6.4}{\nano\second} acquisition interval, these symbol intervals contain at least three and six measured output samples, respectively.
For each target $y_n$, the sample length $B$ specifies the number of consecutive acquired output samples retained from each measured response.
At $B=3$, the sample window contains the first three output samples assigned to the corresponding input-symbol interval.
The third sample defines the readout time, and longer sample windows end at this same point and extend backward into preceding-symbol intervals.
No output sample acquired after this readout time is included.
The evaluated sample lengths are selected according to their maximum symbol depths through the mapping defined below.
The values are $B=3$, 9, 33, and 39.
Let $C$ denote the number of laser-frequency or output-port responses supplied to the readout, and let $x_{n,b}^{(c)}$ denote the $b$th retained sample from response $c$ associated with target $y_n$.
The optical feature vector is
\begin{equation}
	\mathbf{x}_n^{(B)}=
	\left(
	x_{n,1}^{(1)},\ldots,x_{n,B}^{(1)},
	\ldots,
	x_{n,1}^{(C)},\ldots,x_{n,B}^{(C)}
	\right)^{\mathsf{T}},
\end{equation}
The vector contains $B\times C$ components.
An affine linear readout reconstructs the target as
\begin{equation}
	\widehat{y}_n
	=
	b+\mathbf{w}^{\mathsf{T}}\mathbf{x}_n^{(B)},
\end{equation}
where $b$ is the fitted intercept and $\mathbf{w}$ contains the readout weights.
The normalized mean-squared error is
\begin{equation}
	\mathrm{NMSE}
	=\frac{\displaystyle\sum_n\left(y_n-\widehat{y}_n\right)^2}
	{\displaystyle\sum_n\left(y_n-\overline{y}\right)^2},
\end{equation}
The quantity $\overline{y}$ is the mean target over the samples included in the evaluation.
Lower NMSE values indicate more accurate reconstruction.

The digital reference is constructed directly from the normalized input sequence and does not use measured optical-output samples.
With maximum symbol depth $D$, its feature vector is
\begin{equation}
	\mathbf{d}_n^{(D)}=
	\left(u_n,u_{n-1},\ldots,u_{n-D}\right)^{\mathsf{T}}.
\end{equation}
Thus, $D=0$ supplies only the current symbol, and $D>0$ supplies the preceding $D$ symbols in addition to the current symbol.
The digital reference uses the same affine readout and NMSE, and the resonant and off-resonance optical references use $\mathbf{x}_n^{(B)}$.
The quantities $D$ and $B$ represent different properties: $D$ is a maximum symbol depth, while $B$ is a sample length.
For comparison on a common symbol-depth axis, the maximum symbol depth associated with an optical sample length is
\begin{equation}
	D(B)=\left\lfloor\frac{B\Delta t}{T}\right\rfloor.
\end{equation}
For $T=\qty{40}{\nano\second}$, we select sample lengths $B=3$, 9, 33, and 39 to obtain maximum symbol depths $D=0$, 1, 5, and 6, respectively.
These maximum symbol depths cover the principal variations of the digital-reference curve rather than forming a uniform sweep.
At $B=3$, the three acquired samples occupy a nominal sample-window duration $B\Delta t=\qty{19.2}{\nano\second}$.
These samples provide three time-multiplexed readout features.

Each sequence is partitioned chronologically: the first \qty{80}{\percent} forms the training subset, and the final \qty{20}{\percent} forms the held-out test subset.
The analysis discards the first \num{100} points of each partition as washout.

The linear readout is fitted on the training subset, and the held-out test subset provides the final evaluation.

Output-port and laser-frequency responses are combined separately using procedures analogous to the MC procedures described in \cref{sec:methods-mc-simple-ports,sec:methods-mc-greedy-frequency}.
Output-port subsets are evaluated at fixed laser frequency and mean input power.
Laser-frequency response sets are assembled greedily at fixed output port and mean input power.
The frequency-diverse NARMA5 representations use port 1.
For the NARMA-order comparison, the laser-frequency--power operating point is selected independently for each order and sample length while the five-port set remains fixed.
The greedy algorithm further divides the training subset chronologically into fitting and validation subsets.
The fitting subset precedes the validation subset.
The algorithm fits the readout on the fitting subset and uses the validation subset to select each added response.
The test subset is not used during this selection.

The analysis applies this procedure independently to each of the eight input series.
The reported values and error bars are the mean and standard deviation across these series, respectively.

\section*{Data and code availability statement}
The raw measured data supporting the results of this study are available from the corresponding author upon reasonable request. Other data and the code supporting the results of this study is openly available at the following URL/DOI: \url{https://doi.org/10.5281/zenodo.23284555}.

\section*{Acknowledgments}
A.F. acknowledges funding by the European Union under GA n°101070238-NEUROPULS. A.L. acknowledges funding by the European Union under GA n°101064322-ARIADNE. Views and opinions expressed are however those of the author(s) only and do not necessarily reflect those of the European Union or The European Research Executive Agency. Neither the European Union nor the granting authority can be held responsible for them.

\printbibliography

\section*{Contributions}

A.F. conceived the experiments, prepared the setup, performed the experimental measurements, performed the data analysis and wrote the manuscript. A.L. designed the integrated photonic circuit. A.L., P.B. and L.P supervised the work. All authors contributed to the revision of the manuscript.

\end{document}